\documentclass[12pt,preprint2]{aastex} 
\usepackage{graphicx,emulateapj5,apjfonts}
\usepackage{natbib}
\usepackage{onecolfloat}
\usepackage{epsf}

\shortauthors{Gallazzi et al.}
\shorttitle{Obscured star formation in Abell 901/902}

\newcommand{\peg}{{\sc P\'egase }}
\newcommand{\combo}{{\sc combo-17}}
\newcommand{\spitzer}{{\it Spitzer}}
\newcommand{\dens}{$\rm \delta_N$}

\def\aj{AJ}%
\def\araa{ARA\&A}%
\def\apj{ApJ}%
\def\apjl{ApJ}%
\def\apjs{ApJS}%
\def\aap{A\&A}%
\def\aaps{A\&AS}%
\def\mnras{MNRAS}%
\def\pasp{PASP}%
\def\pasj{PASJ}%
\slugcomment{{\sc ApJ in press} \today }

\begin{document}

\def\head{

\title{Obscured star formation in intermediate-density environments: 
A Spitzer study of the Abell 901/902 supercluster}

\author{Anna Gallazzi\altaffilmark{1}, Eric F.\ Bell\altaffilmark{1}, Christian Wolf\altaffilmark{2}, Meghan E.\ Gray\altaffilmark{3}, Casey Papovich\altaffilmark{4}, 
Marco Barden\altaffilmark{5}, Chien Y. Peng\altaffilmark{6}, Klaus Meisenheimer\altaffilmark{1}, Catherine Heymans\altaffilmark{7}, Eelco van Kampen\altaffilmark{5}, 
Rachel Gilmour\altaffilmark{8}, Michael Balogh\altaffilmark{9}, Daniel H. McIntosh\altaffilmark{10}, David Bacon\altaffilmark{11}, Fabio
D. Barazza\altaffilmark{12}, Asmus B\"ohm\altaffilmark{13}, 
John A.R. Caldwell\altaffilmark{14}, Boris H\"au\ss ler\altaffilmark{3}, Knud Jahnke\altaffilmark{1}, Shardha Jogee\altaffilmark{15}, Kyle Lane\altaffilmark{3}, 
Aday R. Robaina\altaffilmark{1}, Sebastian F. Sanchez\altaffilmark{16}, Andy Taylor\altaffilmark{17}, 
Lutz Wisotzki\altaffilmark{12},Xianzhong Zheng\altaffilmark{18}}

\begin{abstract}
We explore the amount of obscured star-formation as a function of environment in the A901/902
supercluster at $z=0.165$ in conjunction with a field sample drawn from the A901 and CDFS fields, imaged with {\it HST} as part of the
STAGES and GEMS surveys. We combine the
\combo\ near-UV/optical SED with \spitzer\ 24\micron\ photometry to estimate both the unobscured and obscured star formation in
galaxies with $\rm M_\ast > 10^{10}M_\odot$. We find that the star formation activity in massive galaxies 
is suppressed in dense environments, in agreement with previous studies. Yet, nearly 40\% of the star-forming galaxies
have red optical colors at intermediate and high densities. These red systems are not starbursting; they have star formation rates per unit
stellar mass similar to or lower than blue star-forming galaxies. More than half of the red star-forming galaxies have low
IR-to-UV luminosity ratios, relatively high Sersic indices and they are equally abundant at all densities. They might be gradually 
quenching their star-formation, possibly but not necessarily under the influence of gas-removing environmental processes. The other $\ga$40\% of the
red star-forming galaxies have high IR-to-UV luminosity ratios, indicative of high dust obscuration. They have relatively high
specific star formation rates and are more abundant at intermediate densities. Our results indicate that while there is
an overall suppression in the star-forming galaxy fraction with density, the small amount of star formation surviving the
cluster environment is to a large extent obscured, suggesting that environmental interactions trigger a phase of obscured star
formation, before complete quenching. 
\end{abstract}

\keywords{galaxies: general ---  
galaxies: evolution --- galaxies: stellar content --}
}

\twocolumn[\head]

\altaffiltext{1} {Max-Planck-Institut f\"ur Astronomie,
K\"onigstuhl 17, D-69117 Heidelberg, Germany; \texttt{gallazzi@mpia.de}}
\altaffiltext{2} {Department of Physics, Denys Wilkinson Bldg., University
of Oxford, Keble Road, Oxford, OX1 3RH, UK}
\altaffiltext{3} {School of Physics and Astronomy, University of Nottingham,
     Nottingham NG7 2RD, UK} 
\altaffiltext{4} {Department of Physics, Texas A\&M University, College Station, TX 77843 USA}
\altaffiltext{5} {Institute for Astro- and Particle Physics, University of Innsbruck, Technikerstr. 25/8, A-6020 Innsbruck, Austria}
\altaffiltext{6} {NRC Herzberg Institute of Astrophysics, 5071 West Saanich Road, Victoria, Canada V9E 2E7}
\altaffiltext{7} {Department of Physics and Astronomy, University of British Columbia, 6224 Agricultural Road, Vancouver, Canada V6T 1Z1}
\altaffiltext{8} {European Southern Observatory, Alonso de Cordova 3107, Vitacura, Casilla 19001, Santiago 19, Chile}
\altaffiltext{11} {Institute of Cosmology and Gravitation, University of Portsmouth, Hampshire Terrace, Portsmouth PO1 2EG}
\altaffiltext{9} {Department of Physics and Astronomy, University Of Waterloo, Waterloo, Ontario, Canada N2L 3G1}
\altaffiltext{12} {Laboratoire d'Astrophysique, \'Ecole Polytechnique F\'ed\'erale de 
Lausanne (EPFL), Observatoire, CH-1290 Sauverny, Switzerland}
\altaffiltext{13} {Astrophysikalisches Institut Potsdam, An der Sternwarte 16, D-14482 Potsdam, Germany}
\altaffiltext{14} {University of Texas, McDonald Observatory, Fort Davis, TX 79734, USA}
\altaffiltext{15} {Department of Astronomy, University of Texas at Austin, 1 University
    Station, C1400 Austin, TX 78712-0259, USA}
\altaffiltext{10} {Department of Astronomy, University of Massachusetts, 710 North Pleasant
    Street, Amherst, MA 01003, USA}
\altaffiltext{16} {Centro Hispano Aleman de Calar Alto, C/Jesus Durban Remon 2-2, E-04004
    Almeria, Spain}
\altaffiltext{17} {The Scottish Universities Physics Alliance (SUPA), Institute for Astronomy, University of Edinburgh, Blackford Hill, 
    Edinburgh, EH9 3HJ, UK}
\altaffiltext{18} {Purple Mountain Observatory, Chinese Academy of Sciences, Nanjing 210008, PR China}

\section{Introduction}\label{sec:intro}
Much observational evidence gathered so far has established that the environment in which galaxies live plays an important role in shaping
their properties, such as their star formation activity, gas content, and morphology, in the sense that galaxies in regions of high galaxy density
tend to have less ongoing star formation, less cold gas and more bulge-dominated morphology
\citep{oemler74,dressler80,lewis02,gavazzi02,gomez03,balogh04a,kauffmann04,mcintosh04,baldry06}. Yet, a real concern is that most star formation
indicators used to date are based on optical properties and are susceptible to the effects of dust attenuation. Indeed a number of studies using
mid-infrared or radio-derived star formation rates (SFRs) have found evidence for some unexpectedly intense bursts of star formation in
intermediate-density regions \citep[e.g.][]{MO02,coia05,fadda08}. The object of this paper is to use wide-field photometric redshift data, deep
\spitzer\ data and wide-field HST imaging of the $z=0.165$ Abell 901/902 supercluster to explore the incidence of dust-obscured star formation
for low-SFR galaxies: is dust-obscured star formation important even at low SFRs, and how does it vary with environment?

\subsection{Environment, star formation and morphology}\label{sec:intro1}
Historically, the first clear evidence that environment influences galaxy properties is the observed 
predominance of early-type galaxies in low redshift clusters with respect to the field, alongside with a paucity of late-type, emission-line
galaxies \citep[e.g.][]{morgan61,dressler80}. The so-called morphology-density relation appears to be in place already at $z\sim1$, but varies quantitatively with redshift:
between $z\sim0.5$ and the present, the fraction of late-type spirals in intermediate-density regions decreases in favour of the population of
S0 galaxies \citep{dressler97,smith05,postman05}. This has suggested that spiral galaxies evolve into smooth and passive systems such as S0s as they
enter the dense environment of galaxy clusters.

Connected to the morphology-density relation is the decrease of the average SFR, as derived from optical colors or emission lines, with increasing
environmental density \citep[e.g.][]{balogh98,gavazzi02,pimbblet02,gomez03}. Among the two, the relation between color (or stellar age) and environment appears to be
the most fundamental one: at fixed color, morphology shows only a weak residual dependence on environment \citep{blanton05,wolf07}. Moreover, the
link between the morphology-density and the SFR-density relations has significant scatter: not all spirals in clusters appear to be
star-forming, at least on the basis of their optical spectra \citep{poggianti99,goto03}. The question remains whether these spirals are really
passive or they have star formation activity that escapes detection in the optical. Indeed, selection of passive spirals on the basis of their
emission lines can be contaminated by dusty early-type spirals with low level of star formation activity, that could instead be detected using e.g.
mid-infrared colors \citep{wilman08}. 

The SFR-density relation extends to very low local galaxy number densities \citep[e.g.][]{lewis02,gomez03} and dark matter densities
\citep{gray04}, corresponding to the outskirts of clusters and the densities of groups. This suggests that not only the cores of
clusters impact galaxy properties but galaxies may experience significant pre-processing in systems with lower density and lower
velocity dispersion such as groups, before entering the denser and hotter environment of the cluster
\citep[e.g.][]{zabludoff02,fujita04}. 

\subsection{Environmental physical processes}\label{sec:intro2}
Several processes can act on galaxies as they interact with their surrounding environment. The intensity and timescale of individual processes
may also vary with galaxy mass and during the galaxy lifetime as it moves through different density environments \citep[for a
review see][]{boselli06}. The gas content and hence star formation activity of galaxies can be affected by interaction with the intra-cluster
medium (ICM).  The cold gas reservoir can be stripped due to the ram-pressure experienced by galaxies falling at high velocities in the dense
ICM of the cluster \citep{gunngott72,quilis00}. Ram-pressure stripping can lead to fast truncation of star formation and its action can be
recognized from truncated H$\alpha$ profiles \citep{koopmann04}, asymmetric gas distribution and deficiency in the cold HI gas \citep{giovanelli85,cayatte90,solanes01} of
many spiral galaxies in local clusters. It is possible that on the front of compression of the cold gas due to ram-pressure a burst of star
formation is induced \citep[e.g.][]{gavazzi85,gavazzi03}. Another gas-stripping process, that affects star formation on longer timescales (of
few Gyrs) than ram-pressure, is the so-called `strangulation' or `starvation': assuming that galaxies are surrounded by a halo of hot diffuse
gas, this can be removed when galaxies become satellites of larger dark matter halos \citep{larson80,balogh00}. Star formation can continue
consuming the cold disk gas, but will eventually die out for the lack of supply of new fresh gas.

The gas distribution, star formation activity and morphology of galaxies can be altered also via interaction with other galaxies. Mergers
between two equally-massive gas-rich galaxies can lead to the formation of a spheroidal system \citep[e.g.][]{toomre72,barnes88,kauffmann93}. The
merger can trigger an intense burst of star formation \citep[e.g.][]{kennicutt87}, rapidly consuming the cold gas and then exhausting due to
feedback processes \citep{springel05}. Merging and slow galaxy-galaxy encounters are favoured in groups and in the infall region of clusters
\citep[e.g.][]{moss06}. At higher densities, galaxies can be affected by the cumulative effect of several rapid encounters with other cluster
members, a mechanism known as `galaxy harassment' \citep{moore98}. After a transient burst of star formation, galaxy harassment leads to
substantial change in morphology. This mechanism can start to operate at intermediate densities, inducing density fluctuations in the gas
\citep{porter08}.

\subsection{Dust-obscured star formation in dense environments?}\label{sec:intro3}
The net effect of the various mechanisms of interaction of galaxies with environment is an accelerated depletion or exhaustion of the gas
reservoir and hence a suppression of the star formation activity. Many of these mechanisms, however, can lead to a temporary enhancement of
star formation, either due to gas compression (e.g. ram-pressure) or density fluctuations that funnel the gas toward the center triggering
nuclear activity (e.g. tidal interactions). The gas and dust column density is likely to increase during such processes and star formation can
be to a large extent obscured and escape optical detection. Star formation indicators that are not affected by dust attenuation need to be
adopted in order to quantify the occurrence of these obscured star formation episodes.

Already several studies based on observations in the thermal infrared (IR) or in the radio have identified significant populations of
IR-bright or radio-bright galaxies in the outer regions of nearby and intermediate-redshift galaxy clusters
\citep[e.g.][]{smail99,MO02,MO03,best04,coia05}. \cite{MO02} find that up to 20\% of the galaxies in 20 nearby Abell clusters have
centrally-concentrated dust-obscured star formation. These galaxies have different spatial distribution with respect to normal star-forming
galaxies or active galactic nuclei (AGN): they are preferentially found in intermediate-density regions. In the A901/902 cluster at $z=0.165$, \cite{WGM05}
have identified an excess of dusty red galaxies with young stellar populations in the intermediate-density, infalling
region of the cluster. Other studies have identified a population of red star-forming galaxies both in the field \citep{hammer97} and in
clusters \citep{verdugo07}. These galaxies could be mistakenly classified as post-starburst on the basis of their weak emission lines
\citep{poggianti99,bekki01}. It is interesting to notice that populations of red, IR-bright star-forming galaxies are often found in filaments
\citep[e.g.][]{fadda00,fadda08,porter08} and in unvirialized or merging clusters \citep[e.g.][]{MO03,geach06,moran07}. Significant populations
of starburst, IR-bright galaxies have been also found in a dynamically young cluster at $z=0.83$ by \cite{marcillac07}. These systems could in
fact be more abundant at higher redshift \citep{saintonge08} as expected from the increase in cosmic star formation activity. Recently, \cite{elbaz07} have shown
that the detection of these galaxies with the use of dust-independent SFR indicators can even lead to a reversal of the star-formation--density
relation at $z\sim1$. 

In this work we want to explore as a function of local galaxy density the importance in the local Universe of the star formation `hidden' among red
galaxies, that would be missed by optical, dust-sensitive SFR indicators. Uniquely, we wish to push to modest SFRs ($\rm \sim 0.2M_\odot~yr^{-1}$), in
order to constrain the star formation mode of typical (not rare starbursting) systems. There are two key requirements for such a study: 1)
obscuration-free SFR indicators, ideally given by the combination of deep thermal IR and UV, in order to obtain a complete census of the total (obscured
and unobscured) SFR; 2) a long baseline in environmental density covering from the cluster cores to the field in order to quantitatively characterize the
SFR-density relation. We analyse the \combo\ CDFS and A901 fields at $z<0.3$, complementing the UV/optical photometry from \combo\ with {\it Spitzer}
24\micron\ data and with {\it HST} V-band imaging from the Galaxy Evolution from Morphology and SEDs (GEMS) survey and the Space Telescope A901/902
Galaxy Evolution Survey (STAGES). The A901 field is particularly interesting in that it contains the supercluster A901/902 at $z=0.165$, a complex system
with four main substructures probably in the process of accreting or merging, where mechanisms altering the star formation and morphological properties
of galaxies might be favoured \citep[e.g.][]{gray02,gray04,gray08,WGM05,wolf07,heymans08}.

We present the sample and the data in Section~\ref{sec:sample} and describe the derivation of SFR and environmental density in
Sections~\ref{sec:parameters} and~\ref{sec:dens}. After discussing the classification into star-forming and quiescent galaxies in
Section~\ref{sec:fractions}, we explore the dependence on local galaxy density of the fraction of (obscured and
unobscured) star-forming galaxies and their contribution to the total star-formation activity as a function of environment 
(Section~\ref{sec:fractions2}). The properties of red star-forming galaxies, such as their SFR, mass, morphology and dust attenuation, are
compared to those of unobscured star-forming galaxies in Section~\ref{sec:redSF}. We summarize and discuss our results in
Section~\ref{sec:conclusion}. Throughout the paper we assume a cosmology with $\rm \Omega_m=0.3$,
$\rm \Omega_\Lambda=0.7$ and $\rm H_0=70km~s^{-1}~Mpc^{-1}$.

\section{The Data}\label{sec:data}
We describe here the sample analysed and the data available. Based on this, we describe the measurement of derived
parameters such as stellar mass, star formation rate (SFR) and environmental density.
\subsection{The sample and the data}\label{sec:sample}
The sample analysed is drawn from two southern fields, the extended Chandra Deep Field South and the A901 field,
covered in optical by the \combo\ survey \citep{wolf03} and at 24\micron\ by MIPS on board the {\it Spitzer Space
Telescope} \citep{rieke04}. \combo\ has imaged three $34'\times33'$ fields (CDFS, A901, S11) down to $\rm R\sim24$
in 5 broad and 12 medium bands sampling the optical spectral energy distribution (SED) from 3500 to 9300\AA. The
17-passband photometry in conjunction with a library of galaxy, star and AGN template spectra has allowed object
classification and redshift assignment for 99\% of the objects, with a redshift accuracy of typically $\delta
z/(1+z)\sim 0.02$.

\spitzer\ has imaged at 24\micron\ a field of $1\deg \times 0.5\deg$ around CDFS as part of the MIPS Guaranteed Time Observations (GTOs) and an
equally-sized field around the Abell 901/902 supercluster (A901 field) as part of \spitzer\ GO-3294 (PI: Bell). The data have been acquired in a scan-map
mode with individual exposures of 10~s. In CDFS, the 24\micron\ data reach a $5\sigma$ depth of 83$\mu$Jy \citep[see][for a technical description of source
detection and photometry]{papovich04}. In A901, the same exposure time reached a 5$\sigma$ depth of 97$\mu$Jy, owing to the high contribution of zodiacal
light at its near-ecliptic position. In what follows, we use both catalogues to 83$\mu$Jy (5$\sigma$ and 4$\sigma$ for CDFS and A901, respectively), noting
that our conclusions are little affected if we adopt brighter limits for sample selection. The 24\micron\ sources have been matched to galaxies with a
photometric redshift estimate in the \combo\ catalogue, adopting a 1'' matching radius. We omit sources within 4' of the bright M8 Mira variable IRAS
09540-0946 to reduce contamination from spurious sources in the wings of its PSF.

The A901 \combo\ field hosts the cluster complex A901/902 composed by the substructures A901a, A901b, A902 and the SW group at a
redshift of $z=0.165$ within a projected area of $\rm 5\times5~Mpc^2~h^{-2}_{70}$. A quarter square degree field centered on the
A901/902 supercluster has been imaged in the filter F606W with the $HST$ Advanced Camera for Surveys (ACS) producing a 80
orbit mosaic, as part of the STAGES survey \citep{gray08}. An area of 800 square arcminutes centered on the extended CDFS
has also been imaged with $HST$ ACS in the F606W and F850LP filters, as part of the GEMS program \citep{rix04}. In the GEMS
survey object detection was carried out using the SExtractor software \citep{sextractor} in a dual configuration that optimizes
deblending and detection threshold \citep{caldwell08}. As described in \cite{gray08}, a similar strategy for source detection
has been adopted in the STAGES survey. Both GEMS and STAGES imaging data have been processed using the pipeline GALAPAGOS
(M. Barden et al. 2009, in prep.), which performs profile fitting and extract Sersic indices (that we then use to morphologically characterise
our sample) with the GALFIT fitting code \citep{Peng02}. 

X-ray data are also available for both the CDFS and the A901 field. X-ray data for the CDFS are available from the $\sim1$Ms {\it Chandra} point
source catalogue published by \cite{alexander03}. The A901 field has been imaged by XMM with a 90~ks exposure and the catalogue is presented in
\cite{gilmour07}. We use the X-ray information to identify possible AGN contribution among star-forming galaxies. To account for the different
sensitivity of {\it Chandra} and XMM, we consider only sources with full band flux $> 1.8\times10^{-15} \rm erg~cm^{-2}~s^{-1}$, the faintest flux
reached in the A901 field.

In this work we wish to study the dependence on environment of the star formation properties of low-redshift galaxies. To this
purpose, we define a sample of galaxies in the redshift range $0.05<z<0.3$ from the CDFS and A901 fields (limited to the areas
covered completely by \spitzer\ and \combo), down to an absolute magnitude of $\rm M_V<-18$ (limited to those objects
classified as galaxies by \combo). The sample peaks at an apparent magnitude of $\rm m_R \sim21$, covering the range $\rm
18\la m_R \la 23$, with a (magnitude-dependent) redshift accuracy of $\sigma_z \la 0.02$ for the majority of the galaxies,
with a tail up to 0.05 \citep{wolf04,WGM05}. The total sample comprises 1865 galaxies (1390 in the A901 field and 475 in the
CDFS), of which 601 have a detection at $\rm 24\micron$ above the $\rm 5\sigma$ level.

We will sometimes refer to `cluster' and `field' sample. The `cluster' sample is defined following \cite{WGM05}, i.e. galaxies in the A901 field with
redshift $0.155<z<0.185$, and it includes 647 galaxies. With this selection of the bright-end of the cluster population,  the completeness reaches about
the 92\% level down to a magnitude of $\rm R\sim23$, but the contamination also rises to 40\% (while it keeps below 20\% for magnitudes brighter than
$\rm R=22$). The `field' sample is defined on the A901 field in the redshift ranges $0.05<z<0.125$ and $0.215<z<0.3$, and on the CDFS over the entire
redshift range $0.05<z<0.3$, with a total of 981 galaxies.

\subsection{Stellar mass and star formation rate}\label{sec:parameters}
Stellar mass estimates have been derived as outlined in \cite{borch06}, using a set of template SEDs generated with the
\peg code, based on a library of three-component model star formation histories (SFHs), devised in such a way to
reproduce the sequence of UV-optical template spectra collected by \cite{kinney96}. The best-fitting SED, and hence
stellar mass-to-light ratio ($\rm M_\ast/L$), is obtained comparing the model colors with the observed ones. Stellar
masses were derived adopting a \cite{kroupa93} initial mass function (IMF). Adopting a \cite{kroupa01} or
\cite{chabrier03} IMF would yield differences in stellar mass of less than 10\%. Random errors amount to $\la 0.3$~dex on a
galaxy-by-galaxy basis, while systematic uncertainties are typically of 0.1~dex for old stellar populations and up to
0.5~dex for galaxies with strong bursts \citep[see also][]{bell07}.

The best indicator of the galaxy SFR combines the bolometric IR luminosity, assuming that it represents the bolometric luminosity of totally
obscured young stars, and total UV luminosity or recombination lines such as H$\alpha$ that trace instead the emission from unobscured young
stars, thus giving a complete census of the luminosity emitted by young stars in a galaxy \citep[e.g.][]{bell03a,calzetti07}. The infrared
data, combined with the NUV-optical \combo\ SED allows us to use such a SFR indicator. For this, we need first to estimate total UV and IR
luminosities from monochromatic information. 

To measure the total IR flux ideally we would need measurements at longer wavelengths \citep[e.g.][]{helou88,dale02}. We only have data in the
24\micron\ MIPS passband which provides us with luminosities at rest-frame wavelength $\sim 23-18.5$\micron\ for the redshift interval
$0.05-0.3$. The monochromatic 12\micron--24\micron\ luminosity correlates well with the total IR luminosity, although it has some residual dependence on
the gas metallicity \citep[e.g.][]{papovich02,relano07,calzetti07}. To convert the 24\micron\ luminosity into total IR luminosity ($8-1000$\micron) we use
the Sbc template of the normal star-forming galaxy VCC 1987 from \cite{devriendt99}. While there is certainly
an intrinsic diversity in infrared spectral shape at given luminosity or stellar mass, this results in a $\la 0.3$dex uncertainty in total
IR luminosity, as inferred using the full range of \cite{devriendt99} templates.

The total UV luminosity ($1216-3000$\AA) is estimated from the luminosity $l_{\nu,2800}$ in the \combo\ synthetic band centered at $2800$\AA\ as
$L_{UV}=1.5\nu l_{\nu,2800}$. The rest-frame $2800$\AA\ band falls blueward of the observed \combo\ U-band (centered at rest-frame 3650\AA) for
galaxies at $z \la 0.3$. The rest-frame luminosity at 2800\AA\ thus requires an extrapolation of the best-fit model over about 200\AA\ at the average
redshift $z\sim0.2$ of the sample. The factor of 1.5 in the definition of $\rm L_{UV}$ accounts for the UV spectral shape of a 100-Myr old stellar population with constant SFR
\citep{bell05}.

We then translate UV and IR luminosities into SFR estimates following the calibration derived by \cite{bell05} from the \peg stellar population
synthesis code, assuming a 100-Myr old stellar population and a \cite{kroupa01} IMF:
\begin{equation}
{SFR[M_\odot yr^{-1}]=9.8\times10^{-11}(L_{IR}+2.2L_{UV})}\label{eqn1}
\end{equation}
This calibration has been derived by \cite{bell05} under the same assumptions adopted in the calibration of \cite{kennicutt98}; the two calibrations
yield SFRs that agree within $\la30$\%. The factor of 2.2 in front of the $\rm L_{UV}$ term in Equation~\ref{eqn1} accounts for the light emitted by
young stars redward of 3000\AA\ and blueward of 1216\AA. We adopt Equation~\ref{eqn1} to estimate the SFR for all galaxies detected at 24\micron. For
galaxies which have upper limits to the 24\micron\ flux, we omit the IR contribution and consider only the UV-optical emission. This is a rather
conservative approach: the SFR of MIPS-undetected galaxies calculated in this way represents a lower limit to the true SFR. On the other hand, including
the $\rm L_{IR}$ term calculated on the basis of the upper limit flux of $83\mu$Jy would overestimate the true SFRs of undetected galaxies.

We note that the adopted calibration relies on the assumption that the infrared luminosity traces the emission from young stars only. There are few
caveats to this assumption. Nuclear activity can also be responsible for at least part of the IR emission. X-ray data and optical identification of
type-1 QSOs on both CDFS and A901 allow us to identify and exclude many AGNs from the sample, but we cannot exclude some contamination from obscured,
Compton-thick AGNs. \cite{risaliti99} find that among local Seyfert 2 galaxies about 75\% are heavily obscured (with hydrogen column densities $\rm
N_H>10^{23}cm^{-2}$) and $\sim$50\% are Compton-thick ($\rm N_H>10^{24}cm^{-2}$). Among all 24\micron-detected galaxies in our sample only $\sim$3\% are
also X-ray detected. Given the relatively faint limit reached in X-ray ($\rm L_X\ga10^{41}erg~cm^{-2}~s^{-1}$), it is reasonable to assume that we
potentially miss Compton-thick sources. Therefore, we expect only a $\sim$3\% contribution by Compton-thick AGNs. Moreover, the presence of an AGN does
not necessarily imply that it dominates the total infrared luminosity \citep{rowanrobinson05}. Indeed A. R. Robaina et al.(2009, in prep.), based on
\cite{ramos07} data and analysis, estimate that type-2 AGNs contribute only $\sim$26\% of the total IR luminosity of their host galaxy. 

In early-type galaxies circumstellar dust around red giant stars is expected to contribute to the mid-IR flux \citep[see][about the sensitivity of IR
bands to different dust components in early-type galaxies]{temi05,temi07}. Nevertheless the mid-IR in early-type galaxies can detect the presence of
intermediate age stars and small amounts of ongoing star formation \citep{bressan07,young08}. As we discuss in Sec.~\ref{sec:fractions}, the majority of
the early-type red-sequence galaxies are not detected at 24\micron. For those that are detected the SFR derived assuming that their IR luminosity traces
young stellar populations is in any case not sufficient to classify them as star-forming galaxies.

There are some caveats also in the use of UV luminosity as tracer of young stars for early-type galaxies. While UV can help to detect recent episodes of
low-level star formation, it can also be affected by evolved stellar populations \citep[e.g.][]{rogers07}. These mainly contribute to the UV upturn at
1200\AA, i.e. at shorter wavelength than what we use, and therefore should not be a concern for the UV luminosities (and SFRs) derived in this work.  

We thus believe that the caveats mentioned above do not appreciably affect the classification into star-forming and quiescent galaxies used in this work
(see Sec.~\ref{sec:fractions}) and our results. The combination of near-UV and deep 24\micron\ data is indeed a powerful tool to detect unobscured and
obscured star formation not only for starbursting galaxies but also in the regime of normal star-forming galaxies.

\subsection{Environmental density}\label{sec:dens}
The combination of the CDFS and of the A901 field, hosting the A901/902 supercluster, provides us with a large dynamic
range of galaxy environments. As mentioned above, we have a well-defined cluster sample, composed of galaxies within
$\pm0.015$ of the redshift of the cluster down to a magnitude of $\rm M_V<-18$, and a comparison field sample. However,
we wish to characterise the environmental galaxy density in a continuous way, such that it allows us to exploit the long
baseline provided by the two fields.

We estimate the environmental density in a cylinder centered at the position of each galaxy in the sample, and express it in terms
of overdensity with respect to an average redshift-dependent background density. The average background density, $\rho_N$, is
calculated combining the three \combo\ fields, in redshift intervals of width 0.1. In each redshift bin the total number of
galaxies, including all objects classified as `galaxy' down to $\rm R=23.5$ and correcting for completeness,\footnote{Galaxy
completeness maps were estimated from simulations as a function of aperture magnitude, redshift and $U-V$ color \citep[see][for a
detailed discussion]{wolf04}.} is divided by the volume given by the total field area and the redshift depth. For each galaxy in
the sample, the local density is obtained by counting the number $N_{gal}$ of galaxies (down to $\rm R=23.5$, correcting for
completeness), in a cylinder centered at the position of the galaxy of radius 0.25~Mpc and depth given by the photometric redshift error
for that galaxy ($\ge 0.015$), and dividing by the volume $V$ of the cylinder corrected for edge effects. The local number density
is then normalized to the average background density interpolated at the redshift of the galaxy. The local overdensity is then
expressed as:
\begin{equation}
{\delta_N = \frac{N_{gal}}{V~\rho_N} - 1}\label{eqn2}
\end{equation}   
This estimate ranges from $\sim -1$ for very underdense regions, to $0$ for average-density regions up to $>4$ for the
densities characteristic of the cluster.

Because of the relatively large errors associated to photometric redshifts (compared to spectroscopic ones) the galaxy density is effectively measured in
volumes that extend $\rm \ga 80~Mpc$ along the line of sight. In this respect the local density adopted here represents a hybrid between projected density
estimates (which neglect redshift information) and spectroscopic estimates (which smooth over much smaller scales of $\rm \la 8~Mpc$). Thus, local densities
calculated with photometric redshifts are biased toward the cosmic mean and suffer on a galaxy-by-galaxy basis from contamination from low-density
interlopers in high-density regions \citep[see][for a comparison of different density indicators]{cooper05}. To quantify this effect, we have tested the
density measures defined in Equation~\ref{eqn2} against mock galaxy catalogues (containing superclusters similar to A901/902), applying the completeness of
the \combo\ survey. Using Equation~\ref{eqn2}, we have measured on the mock catalogues `observed' overdensities assuming realistic photometric redshift
errors (those achieved with \combo, allowing also for catastrophic errors), and `real' overdensities assuming the real observed redshift (including the
peculiar velocity) and a redshift depth of $3\times10^{-3}$. The `observed' overdensities give a density ranking similar to the `real' overdensities, almost
independently of galaxy luminosity and redshift. However, the magnitude of the `observed' overdensities is almost a factor of 10 lower than the `real'
overdensities, owing to the difference in redshift path used to calculate the overdensity.

\begin{figure}
\epsscale{1}
\plotone{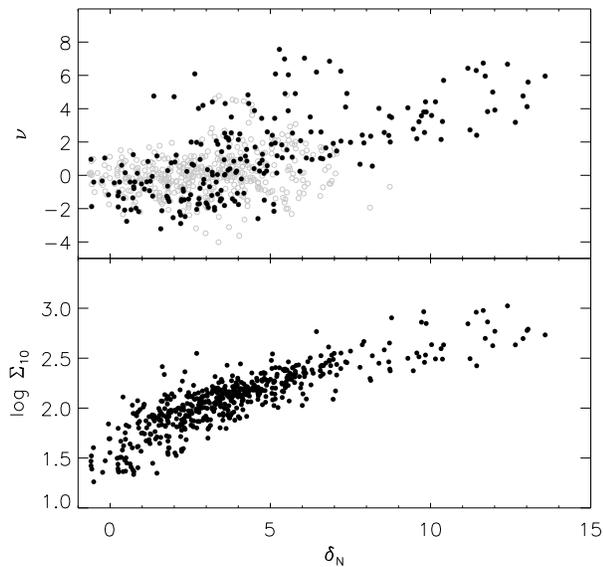}
\caption{The local number overdensity \dens\ used in this work is compared to the surface number density $\rm \Sigma_{10}$ ({\it lower panel}) and to the
surface mass density from weak lensing, as expressed by the parameter $\nu=\kappa/\sigma_n$ ({\it upper panel}). In the upper panel empty grey circles
indicate galaxies for which the dark matter density measure is not reliable because dominated by noise or systematics (see text). The comparison is
performed only on galaxies in the A901/902 cluster.}\label{fig:comp_dens}
\end{figure}

For galaxies in the A901/902 cluster, we could also compare our density estimates to other independent density estimators.
Specifically, we compared with the projected galaxy density $\Sigma_{10}$ as defined by \cite{wolf07}, which measures the number
density of galaxies in an adaptive aperture of radius given by the average of the distance to the $9^{th}$ and $10^{th}$ nearest
neighbour. The lower panel of Fig.~\ref{fig:comp_dens} shows a good correlation between $\Sigma_{10}$ and the galaxy overdensity
\dens\ measured in a fixed aperture. 

The upper panel of Fig.~\ref{fig:comp_dens} compares \dens\ with a measure of the total surface mass density from a weak lensing analysis of the {\it
HST} STAGES data \citep{heymans08}. In this analysis \cite{heymans08} present a pixelated map of the smoothed projected dark matter surface mass
density $\kappa$ of the A901/902 cluster along with noise $\sigma_n$ and systematic error maps $B$, in order to assess the reliability of each
feature. Following \cite{VW00} we define a lensing density measure $\nu = \kappa / \sigma_n$ for the pixel region around each galaxy that corresponds
to $\rm \sim 20\times 20 kpc^2$. For $\nu >> 1$ we can calculate a corresponding mass estimate, following equation 4 in \cite{heymans08}, where a galaxy
with a lensing density measure $\nu = 4$, for example, is enclosed in a local dark matter mass of $\rm M(<20 kpc) = 1 \times 10^ {11} M_\odot$.  For $\nu<1$
we enter a low to underdense regime, with the most negative regions showing the location of voids \citep{JainVW,Miyazaki}. 

\begin{figure}
\epsscale{0.9}
\plotone{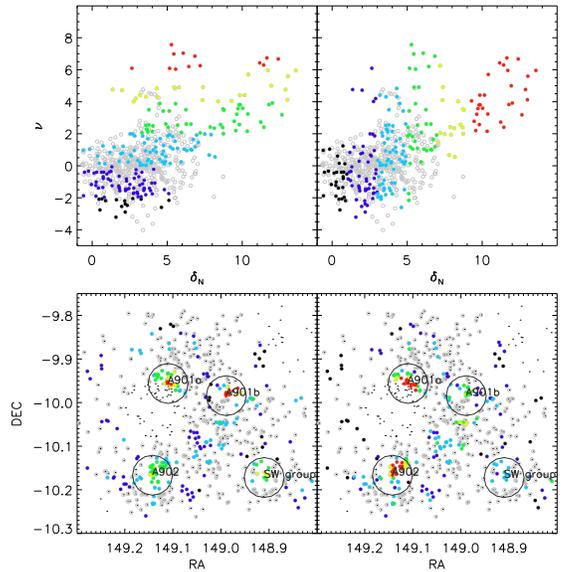}
\caption{{\it Lower panels}: position on the sky of the A901/902 cluster galaxies, color-coded according to their environmental density, expressed either
as dark matter density $\nu$ ({\it left}) or as galaxy number overdensity \dens\ ({\it right}). The small dots refer to all \combo\ $\rm M_V<-18$
galaxies in the A901/902 cluster. The sample analysed is limited to galaxies covered completely by \combo\ and \spitzer. Grey empty circles are galaxies
with an unreliable $\nu$ measures. The galaxies of interest in this comparison, i.e. those with reliable $\nu$ values, are represented with filled
colored circles. {\it Upper panels}: relation between $\nu$ and \dens. The different colors indicate the density ranges in which galaxies are
sorted in the corresponding lower panels.}\label{fig:comp_dens_sky}
\end{figure}

In this paper we are particularly interested in the low to intermediate density regions of the A901/902 cluster. Unfortunately for the weak lensing
analysis however, it is these lower density regions where systematic errors become important. We therefore introduce a selection criteria, following
\cite{heymans08}, that the lensing density estimate $\nu$ is deemed reliable if the systematic error $B$ is either comparable to the noise $\sigma_n$ or
less than half the amplitude of the signal $\kappa$. Fig.~\ref{fig:comp_dens} shows unreliable measures as open points. Comparing the reliable
lensing density measurements $\nu$ (filled points) with \dens\ shows a good correlation between these two environment variables. Taking only those
galaxies with a reliable lensing measure, we show, in the lower-left panel of Fig.~\ref{fig:comp_dens_sky}, the position in the sky of the A901/902
cluster galaxies, color-coded according to their $\nu$ value as indicated in the upper-left panel. The corresponding right-hand panels refer to the
galaxy overdensity \dens. We note in particular that the two dark matter peaks corresponding to the A901a and A902 cores are also identified as peaks in
the galaxy distribution. Galaxies in these regions follow the main relation between $\nu$ and \dens\ shown in the upper panels of
Fig.~\ref{fig:comp_dens}. The A901b core and SW group are instead associated to a lower galaxy density and show a larger spread to higher $\nu$ values at fixed
\dens\ that is not completely explained by larger errors on $\nu$. 

Whilst weak gravitational lensing techniques can provide a direct measure of the total matter density, this environment variable is
integrated along the line of sight with contributions from mass at all redshifts. In the case of the A901/902 cluster it is a reasonable approximation to
place all the measured mass at the redshift of A901/902 as shown by \cite{heymans08} who find that the mass of this supercluster is significantly larger
than the known galaxy groups and the CBI cluster behind A902 \citep{taylor04}. However in the case of the CDFS field, mass is distributed fairly equally
along the line of sight at relatively low density. It would therefore be very difficult to obtain a local matter density measure for this field from a
weak lensing analysis even with the {\it HST} imaging that exists \citep[see][]{heymans05}. For this reason we favour using $\delta_N$ as it permits local
density measurements in both the field and cluster environments.

\begin{figure}
\epsscale{1}
\plotone{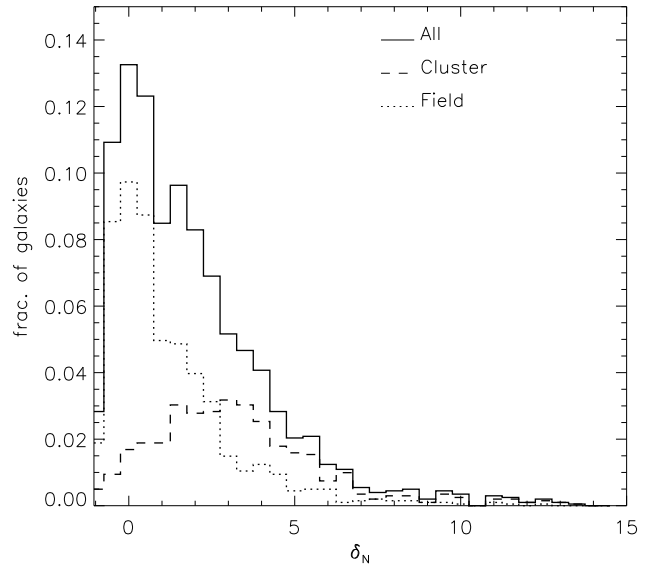}
\caption{Distribution in local density \dens\ for the sample as a whole (solid line), and then split into `cluster'
(dashed line) and `field' galaxies (dotted line).}\label{fig:distr_dens}
\end{figure}

We note that the large redshift depth assumed in the density measure affects in particular the cluster sample, for which one would expect overdensities
higher by about an order of magnitude. In what follows, however, we keep also for cluster galaxies the overdensities estimated over a depth set by the
photometric redshift error, since we want to study cluster and field galaxies simultaneously with a consistent density measure. The distribution in density
\dens\ for the sample as a whole is shown in Fig.~\ref{fig:distr_dens}. The dashed and dotted lines distinguish cluster galaxies from the field sample. As
expected the field sample is concentrated in environments with density similar to or below the average background density. Cluster galaxies instead dominate
at densities above \dens$\sim2$.   


\section{Results}\label{sec:results}
We now describe the classification of galaxies on the basis of their star formation rate and optical color, that we will use throughout the paper
(Sec.~\ref{sec:fractions}). Unless otherwise specified, the terms `red star-forming' and `obscured star-forming' used in the text refer to the same class
of galaxies. We then investigate how the fraction of star-forming galaxies depends on galaxy environment, with particular attention to the extent of star
formation `hidden' among red-sequence galaxies (Sec.~\ref{sec:fractions2}). In Sec.~\ref{sec:redSF} we analyse the star formation properties, morphology,
and dust attenuation of red star-forming galaxies, as opposed to quiescent ellipticals and blue-cloud galaxies, as a function of environment.
\subsection{Galaxy classes}\label{sec:fractions}
\begin{figure}
\epsscale{1}
\plotone{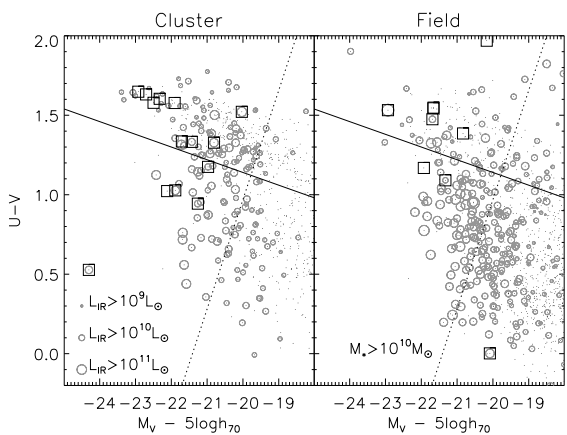}
\caption{Rest-frame color-magnitude diagram for cluster galaxies ({\it left}) and for field galaxies ({\it right}). Grey
circles represent galaxies detected at 24\micron, with symbol size scaling according to the total IR luminosity. Large 
squares indicate galaxies associated to X-ray sources. There is a significant fraction of 24\micron\ emitting galaxies
(with IR luminosity also in excess of $\rm 10^{10}L_\odot$) populating the cluster red sequence (solid line), in
particular with masses greater than $\rm10^{10}M_\odot$ (dotted line).}\label{fig:cmr_clus_field}
\end{figure}

Fig.~\ref{fig:cmr_clus_field} shows the distribution in the color-magnitude plane of galaxies in the A901/902 cluster (left panel) compared to
galaxies in the field (right panel). The solid line indicates the magnitude-dependent color cut adopted to classify galaxies as red-sequence
(redward of the line) or blue-cloud galaxies (blueward of the line). The cut is set $0.25$~mag blueward of the color-magnitude relation fitted
by \cite{bell04} on the combined A901+CDFS fields at $0.2<z<0.3$. Although the exact fraction of blue/red galaxies depends on the chosen color
cut, it makes a little difference as long as the cut lies in the `gap' between the `blue' and the `red' peaks of the color distribution.
Grey circles represent galaxies detected at 24\micron, with symbol size scaling according to their total IR luminosity. While we are not
surprised to find a large number of 24\micron-emitting galaxies in the blue cloud, especially in the field, it is also noticeable a significant
contamination of the cluster red sequence by IR-luminous galaxies. A fraction of the IR luminosity may come from AGNs, although we notice that
only a small number of IR-luminous red-sequence galaxies are identified as X-ray sources (large squares).
We cannot exclude some contamination by obscured, Compton-thick AGNs, but we believe this is only a few percent (see Section~\ref{sec:parameters}).
Fig.~\ref{fig:cmr_clus_field} illustrates that 24\micron\ information allows us to reveal a significant number of red-sequence galaxies with
infrared luminosity in excess of $\rm 10^{10}L_\odot$, witnessing to a large extent ongoing star formation activity onto the red sequence, that
would be otherwise undetected (or at least underestimated) because obscured by dust. 

Before exploring the properties of red IR-luminous galaxies and their importance in terms of the total star formation budget as a function of
environment, we define our classification into quiescent and star-forming galaxies, further distinguished into red and blue. We concentrate on
galaxies more massive than $\rm 10^{10}M_\odot$, thus sampling the high-mass end of the mass function above which the red-sequence completeness
is guaranteed up to redshift 0.3 \citep{borch06}. We set a threshold in specific SFR of $\rm \log (SFR/M_\ast) = -10.7$, which corresponds to a
level of star formation of $\rm 0.2~M_\odot~yr^{-1}$ at the mass limit. We thus define galaxies as `quiescent' or `star-forming' depending on
whether their specific SFR is below or above this level, respectively. We then separate `red SF' and `blue SF' galaxies according to the
magnitude-dependent red-sequence cut shown in Fig.~\ref{fig:cmr_clus_field}. 

\begin{figure}
\epsscale{0.9}
\plotone{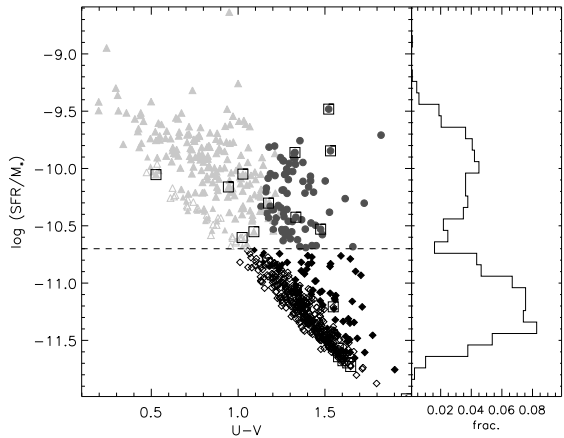}
\caption{{\it Left panel}: Specific SFR against optical color for the massive galaxies ($\rm M_\ast > 10^{10}M_\odot$) in the sample. The dashed line indicates the
level of specific SFR below which galaxies are classified as `quiescent' (black diamonds). Above $\rm \log~ (SFR/M_\ast) = -10.7$ `star-forming'
galaxies are then separated into red-sequence (dark grey circles) and blue-cloud galaxies (light grey triangles). Most of the massive star-forming
galaxies have detectable emission at 24\micron\ (highlighted by filled symbols). X-ray sources are indicated with large squares. The apparent gap
in specific SFR is due to the fact that we use only the UV/optical term in Equation~\ref{eqn1} to estimate SFR for galaxies not detected at
24\micron\ (see text). {\it Right panel}: Distribution in specific SFR for the whole sample of massive galaxies. The bimodal distribution motivates the choice of a
cut in specific SFR at $\rm \log~ (SFR/M_\ast) = -10.7$.}\label{fig:selection}
\end{figure}

The choice of the specific SFR limit is justified by the fact that the distribution of the massive galaxies in the sample in specific SFR (as measured in
Equation~\ref{eqn1}) is bimodal and the two peaks separate at a value of $\sim-10.7$, which is also very close to the mean value of specific SFR for this
sample. This is clearly shown in the right-hand panel of Fig.~\ref{fig:selection}. As discussed in Sec.~\ref{sec:parameters}, for galaxies with IR flux
below the upper limit of 83$\mu$Jy we estimate SFR only from their UV luminosity.  If we included the IR term also for these galaxies the distribution
would no longer be bimodal, and the mean value of specific SFR would be  $\rm \log (SFR/M_\ast) \sim -10.6$. We decide to keep the conservative approach
of using the lower limit SFR for galaxies not detected at 24\micron, however we will mention when relevant how the results would change if we used
instead the upper limit SFR (i.e. adopting the 24\micron\ upper limit flux of 83$\mu$Jy to estimate $\rm L_{IR}$ for non-detections).

Figure~\ref{fig:selection} (left panel) describes our classification for the 689 massive galaxies in the sample, showing their distribution in specific SFR
against the rest-frame $U-V$ color. Quiescent galaxies (below $\rm \log (SFR/M_\ast) = -10.7$, dashed line) are shown as black diamonds and almost all of
them belong to the red sequence. Star-forming galaxies (above the dashed line) are distinguished into blue-cloud galaxies (light grey triangles) and
red-sequence galaxies (dark grey circles). About 60\% of the sample is classified as quiescent (406 galaxies), the remaining is divided into 77 red
SF and 206 blue SF galaxies. Galaxies that have a detection at 24\micron\ are highlighted with filled symbols. We note that all the red SF galaxies
have 24\micron\ detection, while 13\% of the blue SF galaxies are not MIPS detected (their UV-based SFR is thus more properly a lower limit to
the total SFR). Among the quiescent galaxies, 81 have detectable IR emission. Few of the 24\micron-detected quiescent galaxies are assigned a
specific SFR higher than expected on the basis of their color, but it is not clear whether the IR emission in these cases is truly indicative of low level
of star formation or rather comes from circumstellar dust in red giant stars \citep[but see][]{temi07,temi08} or from an AGN (although none of
these galaxies is associated to an X-ray source, as shown by the large squares). In any case, even assuming that the IR emission in these
galaxies is associated to young stars it is not enough to classify them as star-forming. 

It is worth mentioning that the location of galaxies in the specific SFR versus $U-V$ plane is independent of environment, with only the relative
importance of blue SF/red SF/quiescent galaxies changing with environment, as we discuss in Fig.~\ref{fig:color_mass_frac} below.

The apparent gap in specific SFR in Fig.~\ref{fig:selection} between IR-detected and IR-undetected galaxies is due to the drop of the $\rm
L_{IR}$ term in Equation~\ref{eqn1} in the latter case. Adopting an IR luminosity for IR-undetected galaxies given by the upper limit flux of
83$\mu$Jy would increase the specific SFR of these galaxies and fill in the gap somewhat. While this would have a small effect on the number of
blue star-forming galaxies (because their SFRs are already above the threshold even when not detected at 24\micron), the number of red-sequence
galaxies classified as star-forming would increase at the expense of quiescent galaxies. More quantitatively, adopting the upper limits on SFR
and a specific SFR threshold of $\rm \log (SFR/M_\ast) = -10.7$ (as in our default case) or $-10.6$ (the mean value for the `upper limit'
specific SFRs), the number of red SF galaxies would increase to 170 or 126, respectively, while the number of quiescent galaxies would decrease
to 299 or 347, respectively (note that in this case the selection would be more sensitive to the exact cut in specific SFR adopted).

We certainly expect a number of star-forming galaxies to have colors as red as red-sequence galaxies simply due to inclination effects. We have visually
inspected the STAGES and GEMS V-band images of the red SF galaxies in our sample. We found that 19\% of them appear as edge-on spirals with dust lanes on
the plane of the disc. These galaxies might be classified as blue SF if viewed with a different angle. Another 10\% of the red SF galaxies are inclined
spirals but with irregular structure (also in the dust), so it is not clear what the inclination effects in these cases are. We conclude that undisturbed
edge-on spirals can account for no more than 30\% of the red SF galaxies in our sample. There must be an excess population that accounts for the full
sample of red SF galaxies, either old galaxies with some residual star formation or galaxies with inclination-independent dust obscuration or a
combination of both, as we discuss in Section~\ref{sec:redSF}. 

\subsection{Galaxy fractions versus environmental density}\label{sec:fractions2}
\begin{figure*}
\epsscale{1.4}
\plotone{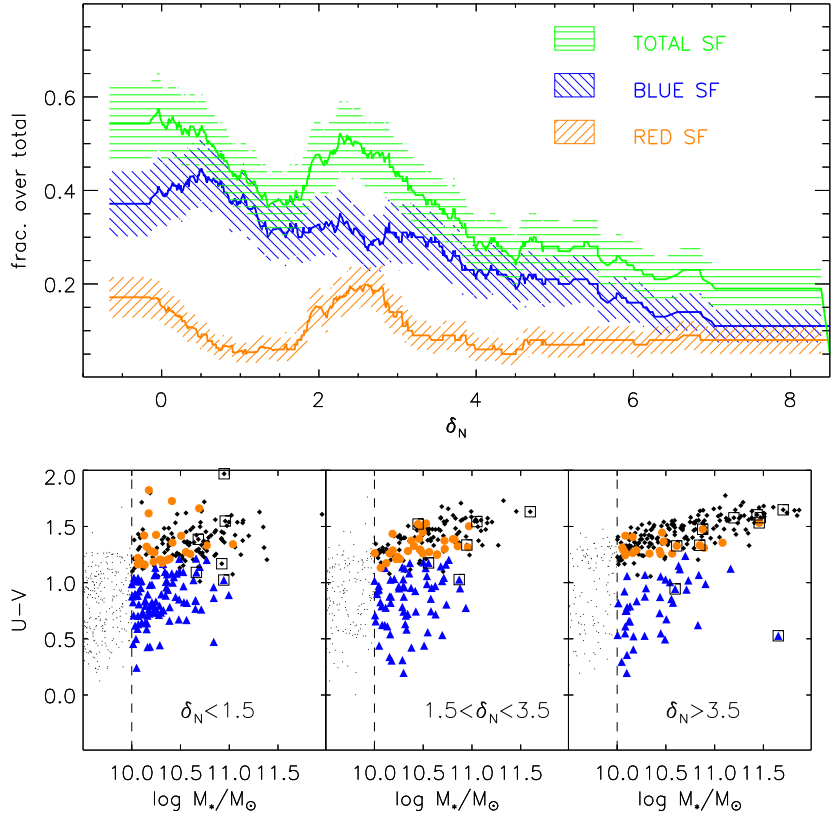}
\caption{{\it Bottom panels}: Rest-frame $U-V$ color against stellar mass for galaxies in low-density environments ({\it left}) compared
to intermediate-density and high-density environments ({\it middle and right}). Above $\rm 10^{10}M_\odot$ (dotted line) we distinguish
quiescent galaxies (black diamonds), blue SF (blue triangles) and red SF galaxies (orange circles). Identified X-ray sources are marked
with a square. {\it Upper panel}: Fraction of unobscured (blue curve) and obscured (orange curve) star-forming galaxies above $\rm
10^{10}M_\odot$ as a function of galaxy number overdensity \dens. The green curve shows the total fraction of star-forming galaxies
(i.e. the sum of the orange and the blue curves). In each case the shaded region encloses the $\pm1\sigma$ Poisson uncertainty. X-ray
sources among star-forming galaxies have been excluded, but including them would make a negligible difference. Contrary to blue SF
galaxies, whose fraction decreases monotonically with \dens, red SF galaxies are found preferentially at intermediate densities where
they constitute $\sim$20\% of the whole population, i.e. only slightly lower than blue SF galaxies.}\label{fig:color_mass_frac}
\end{figure*}

It is not obvious from Fig.~\ref{fig:cmr_clus_field} whether the abundance of 24\micron\ sources `hidden' among red-sequence
galaxies is a feature characteristic of the cluster or whether these sources represent a ubiquitous population. We explore
the possible environmental dependence in Fig.~\ref{fig:color_mass_frac}. Here we do not separate galaxies between `cluster'
and `field', instead we use the continuous definition of environment given by Equation~\ref{eqn2}. The lower panels of
Fig.~\ref{fig:color_mass_frac} show the relation between optical color and stellar mass for galaxies in three disjoint
density regimes, namely low-density environments with \dens$<1.5$, intermediate-density environments with $1.5<$\dens$<3.5$, and
high-density environments with \dens$>3.5$. Different symbols distinguish the three classes of galaxies defined above (galaxies
associated to an X-ray source are indicated with a square): quiescent galaxies (black diamonds), blue SF galaxies (blue triangles) and red
SF galaxies (orange circles). We are particularly interested in the latter class of galaxies, which represents the
class of obscured star-forming galaxies, in comparison to the `unobscured' class, i.e. those galaxies identified as
star-forming also in the optical. Red SF galaxies tend to populate the low-mass end of the red-sequence and their mass range
does not evolve with environment, as opposed to quiescent galaxies. At fixed stellar mass, red SF galaxies are on average
bluer than quiescent galaxies. We will explore these properties in Section~\ref{sec:redSF}.

The upper panel of Fig.~\ref{fig:color_mass_frac} shows the fraction of blue (unobscured) SF and of red (obscured) SF galaxies
among all $\rm M_\ast>10^{10}M_\odot$ galaxies as a function of density. Galaxy fractions are calculated as follows. We first order
galaxies with increasing \dens\ values. For each galaxy we then consider the neighbouring galaxies within a given window in density
($\pm0.5$ of the central value) and calculate the fraction of a given type of galaxies among this subsample. For galaxies in the
first half bin of \dens\ we do not measure fractions but we set their values to the first value actually measured (at
\dens$=-0.5$). The width of the density bin is kept constant until a sufficient number (100) of galaxies fall in that bin. At
higher densities, where the sampling is sparser, we let the bin width vary in order to enclose 100 neighbouring galaxies (50 at
lower densities and 50 at higher densities).\footnote{The density range remains constant up to \dens$\sim4$, it increases to $\pm1$
around \dens$\sim5$. Above \dens$=5$ the density range probed is skewed toward higher densities, but the contamination by
lower-density galaxies does not increase.} When there are not anymore enough neighbouring galaxies we set the fractions to the last
measured values (this happens around a \dens\ of 7). This procedure assures a signal-to-noise of at least 10 with small variation
along the density axis. The shaded regions in the upper panel of Fig.~\ref{fig:color_mass_frac} represent the Poisson
uncertainty in the calculated fractions.  

The blue curve in Fig.~\ref{fig:color_mass_frac} shows the environmental trend of the fraction of unobscured star-forming galaxies.
As expected this fraction decreases significantly with density, from $\sim40$\% at the low densities typical of the field to
$\sim10$\% in the densest environments of the cluster. This trend reflects the well-known decrease in the number of star-forming galaxies
in clusters. When we add the contribution of star-forming galaxies that are on the red
sequence, the overall fraction of star-forming galaxies among massive galaxies (green curve) is increased over the entire density
range covered. What is interesting is that the contribution added by red SF galaxies is not constant with \dens, but produces an enhancement in the
star-forming fraction in particular at densities $1.5\la$\dens$\la 4$.

The orange curve in Fig.~\ref{fig:color_mass_frac} shows the variation with density of the fraction of red SF galaxies. Contrary to blue
SF galaxies, the decrease in the fraction of red SF galaxies with density is not monotonic. At the lowest densities of the field red SF
galaxies represent about 15\% of the total. After an initial decrease from the field toward higher densities, the fraction of red SF
galaxies increases again to values between 15\% and 25\% over the density range $2\la$\dens$\la 3$, and then it settles to a value of
$\la$10\% up to the highest densities of the cluster. The bottom line of Fig.~\ref{fig:color_mass_frac} is that red SF galaxies
represent a non-negligible fraction of the whole galaxy population even at intermediate and high densities. In particular there is an
overabundance of red SF galaxies at intermediate densities where their contribution is comparable to that of blue SF galaxies.

We have checked how the trend of red SF galaxies versus \dens\ would change if we changed the definition of `star-forming' galaxies. If
we included the IR term based on the 24\micron\ upper limit flux in the SFR estimate for MIPS-undetected galaxies, there would be an
overall increase in the fraction of red SF galaxies. This would affect mainly the high-density environments (because of the higher
abundance of red galaxies not detected at 24\micron, likely because genuinely old ellipticals), bringing the red SF fraction between
20\% and 30\% (the exact value depending on the specific SFR cut adopted). Even {\it if} this was correct, it would only strengthen our
main point.

We also checked that the trend in the red SF fraction with density is robust against contamination by edge-on dusty spirals. Even by
removing the $<30$\% contribution by galaxies identified as edge-on spirals (see Sec.~\ref{sec:fractions}), we still detect an
overabundance of red SF galaxies at intermediate densities and the qualitative behaviour with \dens\ does not change.

\begin{figure*}
\plotone{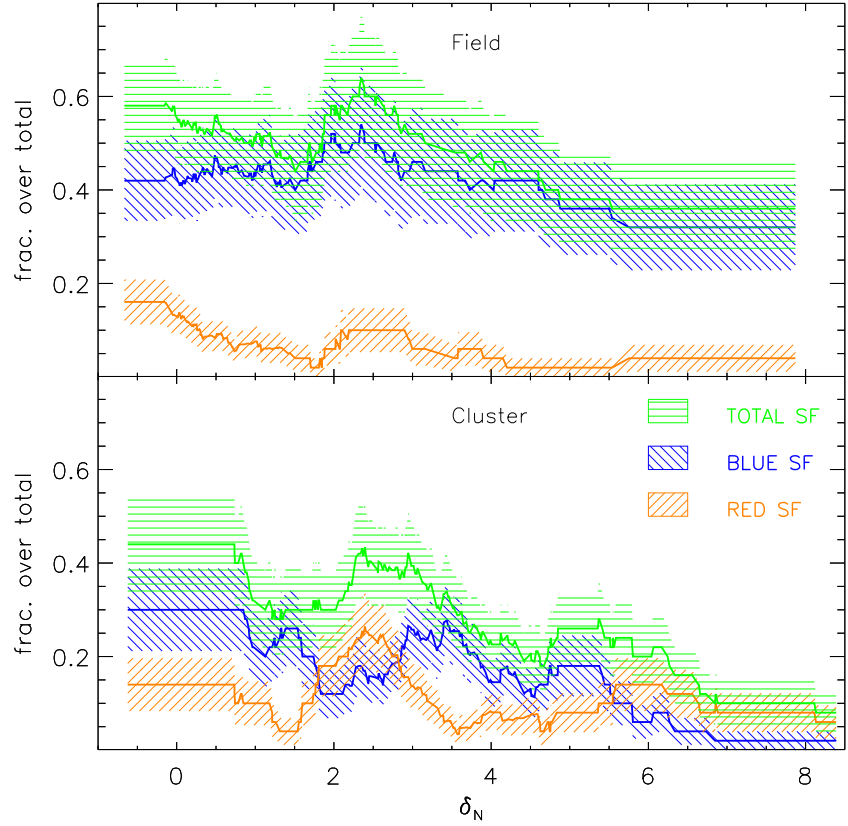}
\caption{Fraction of obscured (orange), unobscured (blue) and all SF galaxies (green) as a function of density \dens\ for the cluster
(lower panel) and for the field sample (upper panel) separately. The relative abundance of red SF galaxies depends on both the local
galaxy number density and the larger-scale environment: the excess of red SF galaxies at intermediate densities is much clearer in
the cluster sample.}\label{fig:frac_clus_field}
\end{figure*}

In Fig.~\ref{fig:frac_clus_field} we show again the fraction of obscured and unobscured SF galaxies as a function of the continuous
density measure \dens\ as in Fig.~\ref{fig:color_mass_frac} but distinguishing galaxies belonging to the A901/902 cluster (lower panel) and
those living in the field (upper panel). In the field sample alone there is only a weak signal of an excess of red SF galaxies at intermediate densities.
The excess found in Fig.~\ref{fig:color_mass_frac} for the sample as a whole is largely driven by cluster galaxies. Fig.~\ref{fig:frac_clus_field} shows
that red SF galaxies are a phenomenon more typical of the cluster environment, where their fraction is comparable to that of
blue SF galaxies. Thus, not only the local galaxy number density but also the larger-scale environment plays a role in shaping the star
formation activity and dust attenuation of galaxies.

\begin{figure*}
\plotone{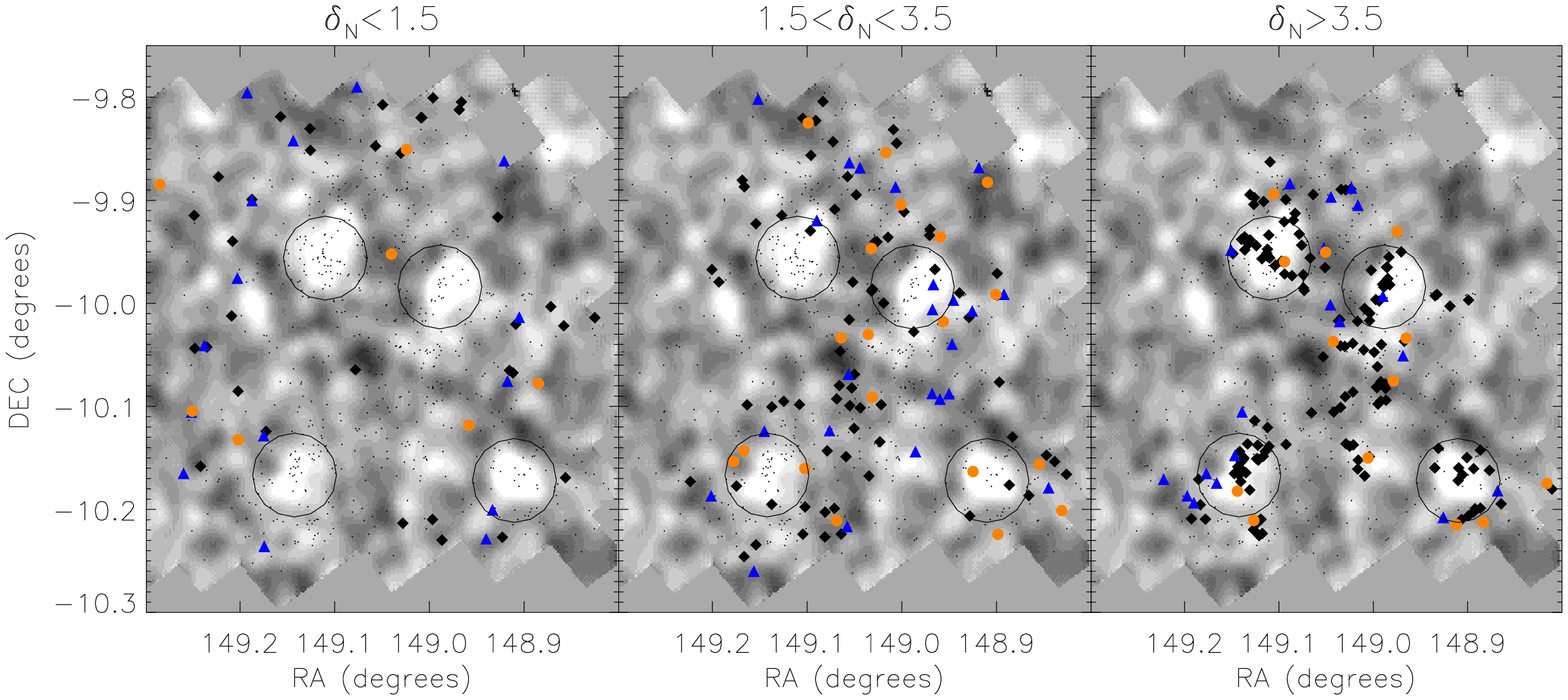}
\caption{Position of quiescent (black diamonds), red SF (orange circles) and blue SF (blue triangles) galaxies in the A901/902 supercluster at
overdensities \dens$<1.5$, $1.5<$\dens$<3.5$, and \dens$>3.5$. The underlying image reproduces the dark matter reconstruction of the
supercluster as derived by \cite{heymans08}, with intensity scaling as the surface mass density $\kappa$. The big circles
indicate the four main supercluster structures (in clockwise order from top-left A901a, A901b, SW group,
A902).}\label{fig:skypos}
\end{figure*}

Fig.~\ref{fig:skypos} illustrates the position on the sky of the cluster red SF galaxies (compared to blue SF and quiescent galaxies) in
the three density ranges of Fig.~\ref{fig:color_mass_frac}. The grey scale shows the dark matter map, as expressed by the surface mass
density $\kappa$, reconstructed by \cite{heymans08} with the STAGES {\it HST} data. Low \dens\ values are typical of the outskirts of
the cluster, mainly populated by blue SF galaxies (left panel). High \dens\ values are instead typical of the four main supercluster
cores and of the filamentary structures connecting them, traced by the quiescent galaxy population (right panel). Red SF galaxies
populate the medium-density regime, the infalling regions around the cluster cores, where episodes of obscured star formation might be
favoured (middle panel). This supports the analysis of \cite{WGM05}, who identified an overabundance in the medium-density regions of
the A901/902 supercluster of dusty, intermediate-age, red galaxies, classified on the basis of their location in optical color-color
diagrams.

\begin{figure}
\epsscale{1}
\plotone{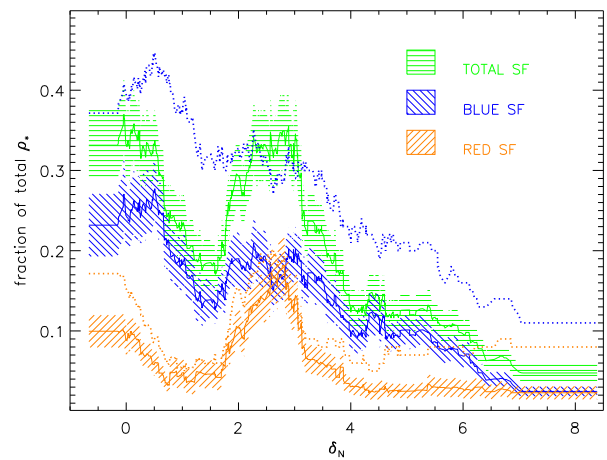}
\caption{Fractional contribution to the total stellar mass density from the sample of massive star-forming galaxies as a whole
(green), and split into blue SF (blue) and red SF (orange). The hatched area around each curve indicates the associated
uncertainty, calculated assuming a 0.3~dex error in stellar mass. For comparison, the blue and orange dotted lines reproduce the
number fractions of blue SF and red SF galaxies, respectively, as shown in Fig.~\ref{fig:color_mass_frac}.}\label{fig:frac_rhom}
\end{figure}

It is also of interest to ask what is the contribution in stellar mass and star formation activity of the different classes of galaxies.
Fig.~\ref{fig:frac_rhom} shows the fraction of stellar mass contributed by $\rm M_\ast>10^{10}M_\odot$ star-forming galaxies (green curve) as a
function of environmental density. As in Fig.~\ref{fig:color_mass_frac} we distinguish star-forming galaxies on the red sequence (orange curve)
and on the blue cloud (blue curve). The stellar mass fraction is calculated in the same way as the number fractions shown in
Fig.~\ref{fig:color_mass_frac}, but weighting each galaxy by its stellar mass. The decline from low to high densities of the stellar mass
fraction contributed by SF galaxies reflects the decline in their number density. The blue and orange dotted lines reproduce the number fraction
of blue SF and red SF galaxies, respectively. At all \dens\ the fraction in mass of star-forming galaxies, either obscured or unobscured, is
lower than the corresponding fraction in number. This comes from the fact that star-forming galaxies are preferentially less massive than
quiescent, elliptical galaxies. The effect becomes stronger at high densities (at least for red SF galaxies), where the mass function of
quiescent early-type red-sequence galaxies extends to higher masses. At low and intermediate densities the
difference between the number and stellar mass fractions is lower for red SF galaxies than for blue SF galaxies, indicating a different stellar
mass distribution of the two classes of galaxies, as we will show in Section~\ref{sec:redSF}. 

\begin{figure}
\epsscale{1}
\plotone{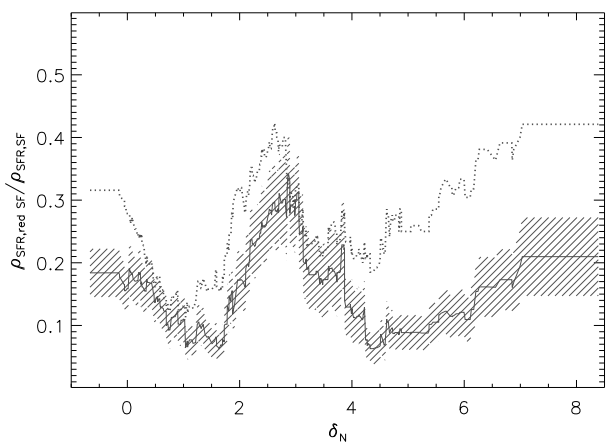}
\caption{Fraction of SFR density contributed by red-sequence SF galaxies. This is compared to their number fraction over the whole SF
population (dotted line). The hatched region represent the uncertainty, calculated assuming a 0.3~dex error on the SFR
estimates.}\label{fig:frac_rhosfr}
\end{figure}

In Fig.~\ref{fig:frac_rhosfr} we investigate the amount of obscured star-formation over the total star formation activity as a function of
environment. This is calculated as the fraction, weighted by SFR, of red SF galaxies over all SF galaxies, and it is shown by the solid curve and
hatched region. For comparison, the dotted curve shows the number fraction of red SF galaxies among all SF galaxies. As expected the majority of
the star formation activity resides in galaxies populating the blue-cloud, independently of environment. Nevertheless, there is a non negligible
contribution, both in number and in total SFR, from obscured star-forming galaxies. In particular there is a clear excess of obscured star
formation at intermediate densities ($2\la$\dens$\la 4$), where red SF galaxies constitute up to 40\% of all SF galaxies and contribute between 25\%
and 35\% of the whole star formation activity at those densities. At higher densities, red SF galaxies still make up $\sim$40\% of the whole SF
class at these densities, but their contribution to the total star formation activity goes down to $\sim$20\%. This suggests a small but
detectable suppression of the SFR of high-density red SF galaxies compared to their intermediate-density counterparts.

\begin{figure}
\epsscale{1}
\plotone{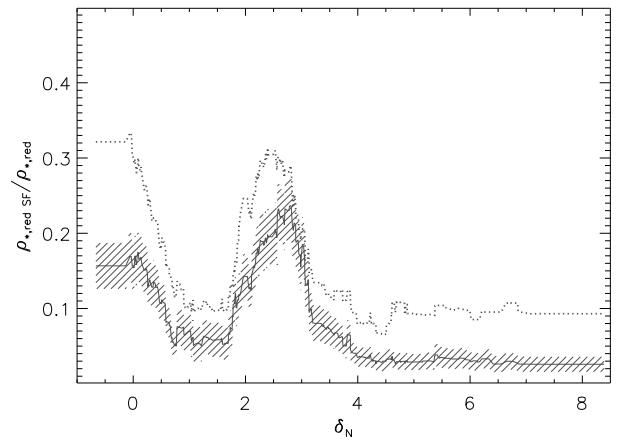}
\caption{Fraction of the stellar mass on the red-sequence contributed by obscured star-forming galaxies with $\rm
M_\ast>10^{10}M_\odot$ (solid curve and hatched region). This is compared to their number fraction over all massive red-sequence
galaxies.}\label{fig:redSF_red}
\end{figure}

Finally, Fig.~\ref{fig:redSF_red} illustrates the amount of contamination on the red sequence from obscured star-forming galaxies. This is
expressed both in terms of the stellar mass fraction (solid line and hatched region) and of the number fraction (dotted line) of
star-forming galaxies among red-sequence galaxies. At low densities, star-forming galaxies contribute roughly 15\% in stellar mass
and 30\% in number to the red sequence. This fraction is in agreement with studies of the mix in morphology and star formation
activity of the `field' red sequence at different redshifts \citep[e.g.][]{franzetti07,cassata07,cassata08}. As expected from the
general decrease in the number of star-forming galaxies in dense environment, the contamination of the red sequence by star-forming
galaxies also decreases with density. However, it reaches values of a few percent only at the highest densities of the cluster,
where the red sequence is highly dominated by quiescent galaxies. At intermediate densities, instead, there is an excess of
(preferentially obscured) star formation, as already discussed in the previous Figures.

\subsection{The properties of red star-forming galaxies}\label{sec:redSF}
In this section we compare the star-formation and morphological properties of red-sequence star-forming galaxies to those of blue-cloud star-forming
and quiescent galaxies. We also investigate any environmental variation of such properties in star-forming galaxies. A follow-up analysis
by \cite{wolf08} based on STAGES data presents environmental trends of the properties of cluster galaxies by distinguishing (visually
classified) morphological types and SED types.

\begin{figure*}
\epsscale{1.4}
\plotone{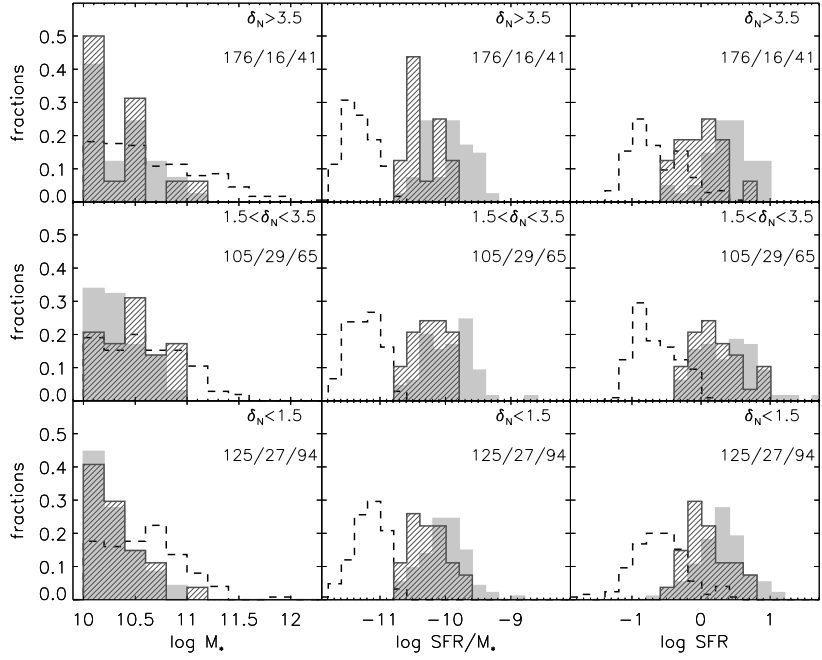}
\caption{{\it From left to right}: distribution in stellar mass, SFR per unit mass and total SFR for different classes of
galaxies in three density regimes (increasing from bottom to top as indicated in each panel). Galaxies are divided into
quiescent (black dashed histograms), blue star forming (grey shaded histograms) and red star-forming galaxies (dark grey hatched histograms). As in
Fig.~\ref{fig:color_mass_frac}, X-ray sources identified among star-forming galaxies are excluded. The total number of quiescent/red-SF/blue-SF galaxies
in each density range is also indicated in each panel. The histograms in each panel are normalized by the number of galaxies in the corresponding class
and in the corresponding density bin.}\label{fig:distr1}
\end{figure*}

Fig.~\ref{fig:distr1} shows the distributions in stellar mass, specific SFR and total SFR of red SF galaxies (hatched histograms) in three density
regimes (\dens$<1.5$, $1.5<$\dens$<3.5$, \dens$>3.5$). In each density range, these distributions are compared to those of blue SF galaxies (grey shaded 
histograms) and quiescent galaxies (dashed histograms). The left panels of Fig.~\ref{fig:distr1} show that, while the stellar mass of quiescent galaxies
clearly increases from low to high densities, the stellar masses of star-forming galaxies hardly vary with density, almost independently of their
obscuration level.  There is however a hint that the mean stellar mass of red SF galaxies at intermediate densities is slightly higher than that of their
low-density and high-density counterparts ($\rm \langle \log (M_\ast/M_\odot)\rangle = 10.47\pm0.05$ compared to $\rm \langle \log
(M_\ast/M_\odot)\rangle = 10.32\pm0.05$ and $\rm \langle \log (M_\ast/M_\odot)\rangle = 10.35\pm0.08$ at low and high densities respectively). The
distribution in stellar mass at intermediate \dens\ compares with that at low \dens\ with a Kolmogorov-Smirnov test probability of 0.03 that the two
distributions are drawn from the same parent distribution. The KS test between the distribution at intermediate \dens\ and at high \dens\ gives a
probability of 0.2. Also, red SF galaxies at intermediate densities are on average more massive than blue SF galaxies at the same densities (which have
$\rm \langle \log (M_\ast/M_\odot)\rangle = 10.34\pm0.03$), with a KS probability of 0.05.

The second and third columns of plots in Fig.~\ref{fig:distr1} show the distributions in specific SFR and total SFR, respectively.
For completeness we also show here the measured SFR of quiescent galaxies, which have by definition specific SFR lower than $\rm
2\times10^{-11} yr^{-1}$. We do not detect any significant variation with environment in the SFR of blue-cloud star-forming galaxies.
Also the specific SFR of red SF galaxies appears independent of environment, but their average SFR at intermediate densities is
slightly higher than at low and high densities ($\rm \langle \log SFR\rangle=0.22\pm0.07$ compared to $\rm \langle \log SFR\rangle=0.07\pm0.06$ and
$\rm \langle \log SFR\rangle=0.01\pm0.08$, respectively), as a consequence of the slightly higher $\rm M_\ast$ discussed above. 

As a general remark, it is interesting to notice that the red, often IR-bright, star-forming galaxies in our sample are not
experiencing a burst of star-formation. They have instead less intense star formation activity compared to
blue-cloud galaxies, independently of environment. Their average specific SFR is from 0.2~dex to 0.3~dex lower than blue SF galaxies
at the 5$\sigma$ level (the KS test on the their specific SFR distributions provides a probability of $0.01$, $0.001$, $0.006$ at
low, intermediate, high densities, respectively). 

\begin{figure*}
\epsscale{1.2}
\plotone{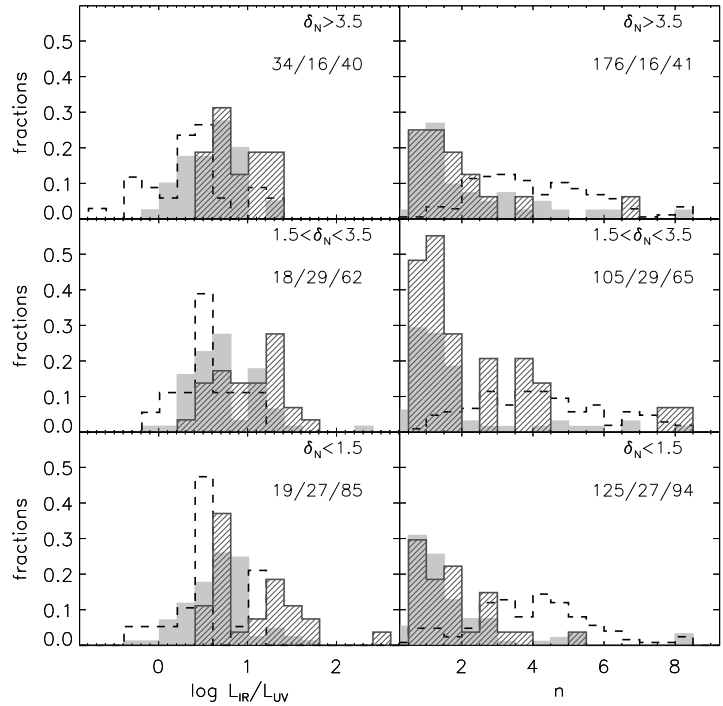}
\caption{Distribution in IR-to-UV luminosity ratio (left) and V-band Sersic index $n$ (right) for quiescent (black dashed histograms), blue SF (grey
shaded histograms) and red SF (dark grey hatched histograms) galaxies in the same density regimes as in Fig.~\ref{fig:distr1}. Note that the distributions
in $\rm L_{IR}/L_{UV}$ are calculated only for galaxies with a detection at 24\micron, not upper limits. This affects only the distribution of quiescent
galaxies, that have a low IR detection rate which decreases with mass. The total number of quiescent/red-SF/blue-SF galaxies
in each density range is also indicated in each panel. The histograms in each panel are normalized by the number of galaxies in the corresponding class
and in the corresponding density bin.}\label{fig:distr2}
\end{figure*}

The right-hand panels of Fig.~\ref{fig:distr2} show the distribution in the V-band Sersic index $n$ for the three classes of galaxies in the three
density regimes. As expected the distribution of star-forming galaxies peaks at low values of $n$ indicating that in all environments star formation
occurs preferentially in disc-dominated systems. This result holds for both unobscured and obscured star-forming galaxies, which have similar
distributions in Sersic index. Also in this case we might witness a difference only at intermediate densities (although with rather low significance),
where red SF galaxies tend to be more bulge-dominated than blue SF galaxies (with a $\langle n\rangle =2.28\pm0.35$ compared to $\langle n\rangle
=1.76\pm0.21$, and a KS probability of 0.12 for the two distributions to be the same).

In this work we have defined star-forming galaxies as `obscured' or `unobscured' only on the basis of their optical color, namely
whether they fall redward or blueward of the red-sequence cut, respectively. The dust attenuation of the UV flux in star-forming
galaxies is often quantified by the ratio of the IR to UV luminosities \citep[e.g.][]{gordon00}.\footnote{The conversion from $\rm L_{IR}/L_{UV}$
to UV attenuation depends on the galaxy star formation activity and the stellar age \citep{cortese08}, but this should not be a concern
for the following discussion. First, the IR luminosity that we infer is based on the luminosity at 24\micron, hence relatively insensitive
to the (typically colder) dust heated by old stars. Second, the red SF galaxies in our sample span only one order of
magnitude in specific SFR, hence the $\rm L_{IR}/L_{UV}$ can at least give us an insight into the {\it relative} dust attenuation among
these galaxies.} We then look at the dust attenuation properties of red SF and blue SF galaxies, as expressed by the IR-to-UV luminosity
ratio, $\rm \log (L_{IR}/L_{UV})$. This is shown in the left panels of Fig.~\ref{fig:distr2} for red and blue SF galaxies in the same
three density regimes as Fig.~\ref{fig:distr1} (for completeness we include also quiescent galaxies; dashed histograms).  The
distributions are shown only for galaxies with a detection at 24\micron\ (for each \dens\ bin the total number of galaxies in each class is indicated
in the panels). All red SF galaxies are detected at 24\micron, while 13\% of the massive blue-cloud galaxies have 24\micron\ flux below
the detection limit. The fraction of blue-cloud galaxies missed is however independent of mass. The 24\micron\ selection in this plot
affects only quiescent galaxies, whose detection rate decreases with mass.
 
As expected, in any density range, red-sequence SF galaxies have on average higher $\rm L_{IR}/L_{UV}$ than blue-cloud galaxies,
indicating a higher level of dust attenuation. The distribution in $\rm \log (L_{IR}/L_{UV})$ differ most significantly at low and
intermediate densities (with a KS probability of 0.003 and 0.002, respectively). In the lowest-density bin, red SF galaxies have an
average $\rm \log (L_{IR}/L_{UV})$ of $1.07\pm0.09$, compared to the $0.68\pm0.04$ of blue SF galaxies. At intermediate densities the
average $\rm \log (L_{IR}/L_{UV})$ of red SF galaxies is $1.04\pm0.07$ compared to $0.73\pm0.05$ of blue SF galaxies. At the highest
densities of the cluster, red SF galaxies still have higher dust attenuation with respect to blue SF galaxies (with a $\rm \langle
\log (L_{IR}/L_{UV})\rangle =0.87\pm0.08$ compared to $0.6\pm0.05$ of blue SF), although the difference between the two distributions
is less significant (with a KS probability of 0.12).

Contrary to the other parameters analysed so far, the IR-to-UV luminosity ratios of red SF galaxies appear to have a roughly bimodal
distribution, with a peak around $\rm \log(L_{IR}/L_{UV})$ values similar to the main population of blue SF galaxies and another peak at
significantly higher values. This is particularly evident at low densities but seems to persist in all environments with varying proportion
between the two groups of red SF galaxies. We can explicitly distinguish red SF galaxies on the basis of their IR-to-UV luminosity ratio,
choosing a cut at $\rm \log(L_{IR}/L_{UV})=1$. By doing so, we find out that {\it low-attenuation} red SF galaxies differ from {\it
high-attenuation} red SF galaxies in their specific SFR, their morphology and their environmental dependence, suggesting that different
evolutionary mechanisms are acting on them.

Low-attenuation red SF galaxies have systematically lower specific SFR than high-attenuation red SF galaxies (with an average $\rm \log
(SFR/M_\ast)$, over all environments, of $-10.43\pm0.03$ compared to $-10.04\pm0.04$ for the latter class). The distributions in specific SFR of
low-attenuation and high-attenuation red SF galaxies differ most significantly at low and intermediate densities (with a KS test probability of
0.004 and 0.002 respectively). Moreover, although with less significance, it is interesting to note that low-attenuation red SF galaxies tend to be
fitted by higher values of Sersic index than high-attenuation red SF galaxies (with a $\langle n\rangle=2.43\pm0.96$ compared to $\langle
n\rangle=1.54\pm0.42$ for the latter class, averaged over all densities). The morphology of both galaxy classes, at least as quantified by $n$, is not a
function of environment.

The fraction of low-attenuation red SF galaxies over the entire population of SF galaxies varies from $11.6\pm3$\% at low densities to
$14.9\pm4$\% at intermediate densities and $17.5\pm6$\% at high densities. There might be a tendency of low-attenuation red SF galaxies becoming
progressively more
frequent in high-density environments, but the errors make these fractions consistent with being independent of environment. On the contrary,
high-attenuation red SF galaxies appear to be more abundant at intermediate densities at about the $2\sigma$ level: at intermediate densities they
represent $16\pm4$\% of all SF galaxies, compared to $10.7\pm3$\% at low densities and $10.5\pm4$\% at high densities. Their stellar mass is also
slightly higher at intermediate densities ($\rm \langle \log (M_\ast/M_\odot)\rangle =10.52\pm0.07$ compared to $10.33\pm0.06$ and $10.28\pm0.1$ at
low and high densities, respectively).

It is worth noting that also among blue SF galaxies there is a subsample of galaxies with $\rm \log (L_{IR}/L_{UV})>1$. By selecting
high-attenuation star-forming galaxies independently of optical color, the same picture emerges in comparison to low-attenuation red SF galaxies as
outlined above. Indeed, the argument of an overabundance at intermediate densities of dust-obscured star formation would be even stronger: the
fraction of $\rm \log (L_{IR}/L_{UV})>1$ SF galaxies among all SF galaxies would be $34\pm7$\% at intermediate \dens, compared to $18\pm4$\% at low
\dens\ and $16\pm6$\% at high \dens.

\begin{figure}
\epsscale{1}
\plotone{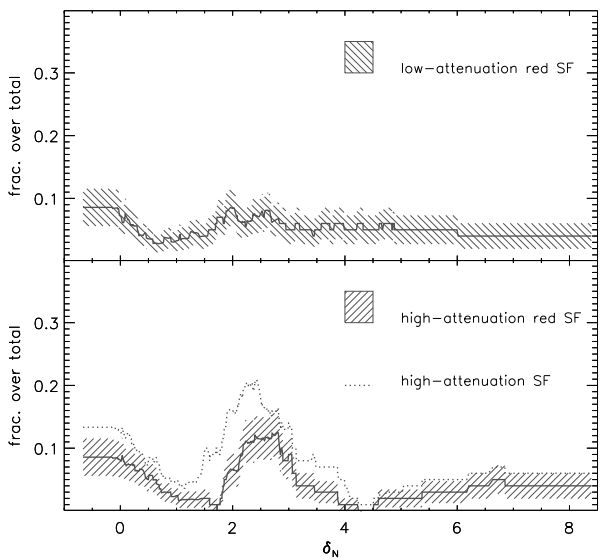}
\caption{{\it Upper panel}: fraction of low-attenuation ($\rm \log(L_{IR}/L_{UV})<1$) red SF galaxies over all galaxies as a function of 
overdensity \dens: they represent on average $\sim$5\% of the whole population, almost independent of environment. {\it Lower panel}: fraction of
high-attenuation ($\rm \log(L_{IR}/L_{UV})>1$) red SF galaxies versus \dens: there is an excess of dust-obscured star formation at intermediate
densities. The same picture emerges if we consider all (blue and red) high-attenuation star-forming galaxies (dotted curve).}\label{fig:fig14}
\end{figure}

This is further illustrated in Fig.~\ref{fig:fig14} which shows the fraction versus the galaxy overdensity \dens\ of red SF galaxies, separated into
low-attenuation (upper panel) and high-attenuation (lower panel). Low-attenuation red SF galaxies constitute on average $\sim$5\% of the whole sample
with no significant dependence on environment. The excess at intermediate densities of red-sequence star-formation identified in
Fig.~\ref{fig:color_mass_frac} is mainly contributed by high-attenuation red SF galaxies, which represent about 12\% of the whole population at
\dens$\sim$2.5. Even excluding the edge-on spirals among high-attenuation red SF galaxies (see Sec.~\ref{sec:fractions}), the trend with \dens\ is still
consistent within the errors with that shown in Fig.~\ref{fig:fig14}. Moreover, dust-obscured star formation could represent up to 20\% of the whole
population at these densities by considering all star-forming galaxies with $\rm \log (L_{IR}/L_{UV})>1$, regardless of their optical color (dotted
curve).

Based on the considerations above, we can say that {\it low-attenuation} red SF galaxies are likely spirals that are gradually quenching
their star formation and appear bulge-dominated because of disc fading. They might resemble the anemic spirals found in local clusters as Coma and
Virgo \citep{vdbergh76,kennicutt83,gavazzi02,gavazzi06}. Some mechanism that removes gas on relatively long timescales {\it could} be responsible
for their transformation toward quiescence. However, given their negligible environmental dependence, internal processes leading to star formation
quenching are equally possible and maybe even sufficient. On the other hand, {\it high-attenuation} red SF galaxies are disc-dominated spirals affected
by some mechanism, particularly efficient at intermediate densities, that triggers obscured episodes of star formation without significantly
changing the morphology, at least on timescales over which star formation is still detectable.


\section{Discussion and conclusions}\label{sec:conclusion}

\subsection{Star formation among red galaxies}
We have combined \combo\ optical data with MIPS 24\micron\ data for a sample of low-redshift ($0.05<z<0.3$) galaxies in the CDFS and
A901 fields with the aim of studying the occurrence of obscured star formation as a function of environment. The 24\micron\
information allows us to recover directly the flux from young stellar populations absorbed and re-emitted by dust, and thus to
trace, in combination with the UV/optical information, the total (unobscured and obscured) star formation activity in galaxies. 
The A901 field is particularly suited from this kind of analysis, not only for the exceptional multiwavelength coverage, but also
because it includes the complex A901/902 supercluster at $z=0.165$ extending over an area of 5$\times$5~Mpc$^2$h$_{70}^{-2}$. The supercluster is
composed of four main substructures, probably in the process of merging. The complex dynamical state of the A901/902 supercluster
potentially makes it an ideal case for identifying galaxies in their process of evolution under the influence of environment. The CDFS,
with the same multiwavelength coverage, offers instead a control sample of field galaxies at similar redshift as the cluster.

In this work we have focused on galaxies with stellar masses larger than $10^{10}M_\odot$, above which the red sequence is complete out to
$z=0.3$ (our limiting redshift). This mass limit roughly corresponds to $0.1\times M^\ast$ over the redshift range $z<0.3$
\citep[]{bell03b,borch06}. We define as star-forming those galaxies with a specific SFR (derived from UV and IR luminosities) above $\rm
2\times 10^{-11} yr^{-1}$. Our focus is on star-forming galaxies populating the red sequence, either because they show low levels of star formation
insufficient to alter the color of the underlying older population or because their star formation activity is highly obscured by dust. Studies
based on the UV and optical emission of galaxies have identified a significant amount of low-level star-formation in low-mass ellipticals
\citep[]{yi05,kaviraj07}, with a hint of a peak in `frosting' activity at group densities \citep[]{rogers07}. Star formation indicators
that are less sensitive to dust attenuation, such as the 24\micron\ emission that we exploit in this work, are instead required to detect
dust-obscured star formation. 

We have studied the abundance of blue and red star-forming galaxies as a function of environment, as expressed by the galaxy number
overdensity in a radius of 0.25~Mpc, focusing on the contribution of star formation on the red sequence compared to
optically-detectable star formation. Our results can be summarized as follows.
\begin{itemize}
\item[-] The overall fraction of star-forming galaxies decreases from $\sim$60\% in
underdense regions to $\sim$20\% in high-density regions. The stellar mass fraction contributed by star-forming galaxies also decreases
going from the field to the cluster cores. The decline is steeper than for the number fraction because, while no significant environmental evolution
in stellar mass occurs for star-forming galaxies, the mass function of quiescent galaxies reaches higher stellar
masses at higher densities.
\item[-] The fraction of blue star-forming galaxies decreases monotonically from $\sim$40\% at low densities to less than 20\% at higher densities. On
the contrary, red SF galaxies do not show a monotonic behaviour as a function of environment. After an initial decline of the red SF fraction
from the field to higher \dens, we identify an overabundance of obscured star formation at intermediate densities, those typical of the outskirts of the
A901/902 supercluster cores. At both intermediate and high densities, red SF galaxies represent 40\% of all star-forming galaxies and contribute
$20-30$\% of the total star formation activity at these densities.
\end{itemize}

To a first order, our results confirm the well-known SFR-density relation \citep[e.g.][]{gavazzi02,lewis02,gomez03,balogh04b,kauffmann04} and
morphology-density relation \citep[e.g.]{dressler80,dressler97,treu03,vdW08}. In addition to this, we find a significant contribution by red-sequence
galaxies, identified as star-forming through their IR emission, to the total star formation activity up to the highest densities of the cluster. This
would be at least partly missed by optical studies. This result is consistent with \cite{WGM05} who already found an enhancement of optically-classified
dusty red galaxies in the medium-density outskirts of the A901/902 supercluster. In this work, supported by deep 24\micron\ data, we directly measures
the amount of star formation going on in these galaxies. 

Our results are also in line with recent studies of clusters at similar redshifts as A901/902 or higher, which have identified a population of IR-bright
galaxies in filaments and the infalling regions of the clusters. \cite{fadda00} found a population of 15\micron-detected galaxies with high
15\micron-to-optical flux ratio suggesting star formation activity in the cluster A1689 at $z=0.18$, in excess with respect to the Virgo and Coma
cluster \citep[see also][]{duc02}. In the cluster A2667 at $z=0.23$ \cite{cortese07} have identified a IR-bright $L^\ast$ spiral galaxy in the process of being transformed by the cluster
environment which triggers an intense burst of star formation. At similar redshift, \cite{fadda08} find two filamentary structures in the outskirts of
the A1763 cluster at $z=0.23$ (probably undergoing accretion events), which are rich in actively star-forming galaxies. \cite{geach06} find an excess of
mid-infrared sources in an unvirialized cluster at $z\sim0.4$, where star formation might be triggered via mergers or interactions between gas-rich
spirals. However, they also note that significant cluster-to-cluster variations are possible: they do not find any significant excess in another cluster
at similar redshift, of similar mass but with a hotter and smoother ICM. Moving to higher redshift, \cite{marcillac07} studied 24\micron\ sources in a
massive, dynamically young, unvirialized cluster at $z=0.83$. They find that IR-detected galaxies tend to lie in the outskirts of the cluster, while they
avoid the merging region. Finally, \cite{elbaz07}, utilizing 24\micron\ imaging in the GOODS fields at redshift $0.8<z<1.2$, have identified for the
first time a reversal of the SFR-density relation observed at lower redshifts. This result has been recently confirmed by \cite{cooper08} with a
spectroscopic analysis using DEEP2 data.

\subsection{Dusty or old?}
The relative abundance of red SF galaxies at intermediate and high densities suggests that they are transforming under the influence
of some environment-related process. What are the star formation activity, morphology and dust attenuation of these red star-forming galaxies?
\begin{itemize}
\item[-] The red SF galaxies in our sample are not in a starburst phase. The few starburst galaxies (with $\rm \log (SFR/M_\ast)>-9.7$,
corresponding to a birthrate parameter $b>1$, assuming a formation redshift of 4) in our sample all populate the blue cloud. We find that red SF galaxies have
similar SFR as blue SF galaxies, and slightly lower specific SFR. While the overall fraction of star-forming galaxies decreases with
density, we do not identify any significant evolution in their level of activity, either obscured or not.  
\item[-] The morphology of star-forming galaxies is not very sensitive to their color. Red SF have similar distribution in Sersic index as
blue SF: they are predominantly disc-dominated. Moreover, the morphology of star-forming galaxies depends little on the environment in which
they live. This suggests that, on average, changes in stellar populations and changes in morphology happen on different timescales, as hinted
at by the fact that color seems to be more sensitive to environment than morphology \citep{blanton05}. The rise of red massive spirals in the
infalling regions of the A901/902 cluster has also been interpreted by \cite{wolf08} as due to SFR decline not accompanied by morphological
change. A two-step scenario in which star-formation is quenched first and morphological transformation follows on longer timescale is also
supported by the analysis of \cite{sanchez07} of the A2218 cluster at $z=0.17$.
\item[-] Red SF galaxies have IR-to-UV luminosity ratios ($\rm L_{IR}/L_{UV}$), a proxy for the level of UV attenuation by dust, on average
higher than blue SF galaxies. The distribution in their IR-to-UV luminosity ratios suggests however the presence of two different
populations, hence possibly two distinct mechanisms affecting star formation activity of red galaxies. Roughly half of the red SF galaxies
in our sample have relatively low $\rm L_{IR}/L_{UV}$, similar to the average value of the bulk of blue SF galaxies, without evolution with
environment. The other half of the red SF galaxies have instead systematically higher $\rm L_{IR}/L_{UV}$. The range in dust attenuation of
this second population becomes narrower at higher densities, suggesting a trend of decreasing attenuation with density.
\end{itemize}

On the basis of the IR properties of red SF galaxies we tentatively distinguish them into two subpopulations. Low-attenuation red SF
galaxies have low specific SFR ($\la10^{-10.3}$) independent of environment. Among star-forming galaxies they tend to have higher Sersic
indices ($\langle n\rangle\sim2.5$). These properties suggest that these galaxies are dominated by rather old stellar populations but have some
residual star formation.
They {\it could} be anemic/gas-deficient spirals gradually suppressing their star formation as a consequence of the removal of their gas reservoir as
they move into higher-density environments \citep{fumagalli_gavazzi08}. Their star formation could be suppressed on relatively long timescales (of few Gyrs) if strangulation of the
hot, diffuse gas occurs while the galaxies enter a more massive halo \citep[e.g.][]{balogh00,vdB08}. The gradual fading of the disc would make the
morphology of these galaxies appear of earlier type. In addition to strangulation, when the density of the surrounding medium becomes sufficiently high,
ram-pressure can act on smaller mass galaxies to remove the remaining gas on the disc and lead to fast quenching
\citep[e.g.][]{gunngott72,quilis00,boselli06b}. However, there is no significant evidence that the relative abundance of low-attenuation red SF galaxies
varies with environment. Therefore, we cannot exclude that these galaxies are suppressing their SF due to internal feedback processes only, without any
additional environmental action required.

The other subpopulation of red SF galaxies have systematically higher $\rm L_{IR}/L_{UV}$, indicative of higher levels of dust attenuation.  They
represent $\ga$40\% of all red SF galaxies in the sample, even after accounting for purely edge-on spirals. By visual inspection of their {\it HST} V-band
images, we can say that the majority of them are spiral galaxies with a bright nucleus or inner bar/disk, suggesting intense star formation activity in
the galaxy core (we cannot exclude AGN contribution in some cases). We also find few cases of interacting galaxies and merger remnants. In comparison to
the low-attenuation red SF class discussed above, they have systematically higher specific SFR and lower values of Sersic index ($\langle
n\rangle\sim1.5$). As opposed to low-attenuation red SF galaxies, they tend to be more abundant at intermediate densities where their stellar mass is
$\sim$50\% and $\sim$70\% higher than at low and high densities, respectively. This suggests that environmental interactions are particularly efficient
in triggering episodes of obscured, often centrally concentrated, star formation in these massive late-type spirals. While we find few cases of
interacting galaxies, violent processes, such as mergers, leading to intense starbursts cannot be the dominant phenomenon. These galaxies are likely more
sensitive to more gentle mechanisms that perturb the distribution of gas inducing star formation (but not a starburst) and at the same time increase the
gas/dust column density. This process should not alter morphology as long as star formation is still detectable. The fact that galaxies undergoing this
phase are preferentially found at intermediate densities and with relatively high stellar masses might indicate longer duration of the dust-obscured
episode of SF for more massive galaxies, which are thus more likely to be caught in this phase than low-mass galaxies. Harassment can act on massive
spirals, funnelling the gas toward the center and leading to a temporary enhancement of star formation \citep{moore98,lake98}. The timescales of this
process could be relatively long if it occurs at group-like densities, rather than in the cluster. Tidal interactions between galaxies at low and
intermediate densities can also produce gas funnelling toward the center \citep{mihos04}.
\\

In summary, we have identified a significant amount of star formation `hidden' among red-sequence galaxies, contributing at least 30\% to the total star
formation activity at intermediate and high densities. The red SF population is composed partly of disc galaxies  dominated by old stellar populations
and with low-level residual star formation, and partly of spirals or irregular galaxies undergoing modest (non-starburst) episodes of dust-obscured star
formation. This means that, while we confirm the general suppression of star formation with increasing environmental density, the small amount of star
formation surviving the cluster happens to a large extent in galaxies either obscured or dominated by old stellar populations. Low-attenuation red SF
galaxies seem to be a ubiquitous population at all densities. Therefore an environmental action is not necessarily required to explain their ongoing
low-level star formation. On the contrary, dusty SF galaxies are relatively more abundant at intermediate densities. They might be experiencing
harassment or tidal interactions with other galaxies, which funnel gas toward the center inducing a (partly or totally obscured) episode of star
formation. Ram-pressure can also be partly responsible for the population of relatively more massive dusty SF galaxies in the cluster: while it is not
effective in removing the gas from the disc in massive galaxies, it could perturb it inducing obscured star formation. The complex dynamical state of the
A901/902 supercluster could favour a combination of different processes producing a temporary enhancement of obscured star formation.

\acknowledgements
A.G. thanks St{\'e}phane Charlot for comments on an early draft and Stefano Zibetti for useful discussions. A.G., E.F.B., A.R.R. and K.J. acknowledge support from
the Deutsche Forschungsgemeinschaft through the Emmy Noether Programme, C.W. from a PPARC Advanced Fellowship, M.E.G. from an Anne McLaren Research
Fellowship, M.B. and E.vK. by the Austrian Science Foundation F.W.F. under grant P18416. C.Y.P. is grateful for support provided through STScI and NRC-HIA
Fellowship programmes. C.H. acknowledges the support of a European Commission Programme Sixth Framework Marie Curie Outgoing International Fellowship under
contract MOIF-CT-2006-21891, and a CITA National fellowship. D.H.M. acknowledges support from the National Aeronautics and Space Administration (NASA) under
LTSA Grant NAG5-13102 issued through the Office of Space Science. A.B. was supported by the DLR (50 OR 0404), S.J. by NASA under LTSA Grant NAG5-13063 and
NSF under AST-0607748, S.F.S. by the Spanish MEC grants AYA2005-09413-C02-02 and the PAI of the Junta de Andaluc{\'i}a as research group FQM322. Support for
STAGES was provided by NASA through GO-10395 from STScI operated by AURA under NAS5-26555.

\bibliographystyle{apj}

\end{document}